\documentclass[12pt]{article}
\usepackage{amsfonts}
\usepackage{amsmath,amssymb}
\usepackage{graphicx}

\baselineskip %
\advance %
\textheight by %
\topskip %
\makeatletter \@addtoreset{equation}{section}
\def\theequation{\thesection.\arabic{equation}}
\makeatother
\def\Z{\mathbb Z}

\def\R{\mathbb R}

\def\be{\begin{equation}}
\def\ee{\end{equation}}

\def\bea{\begin{eqnarray}}

\def\obj(#1)(#2)#3#4#5{%
  \psline[arrows={[-]}, linestyle=dashed, dash=0.0 1,dashadjust=false](#1)(#2)%
  \uput{0.4}[90](#1){#3}%
  \uput{0.4}[-90](#1){#4}\uput{0.4}[-90](#2){#5}%
}
\def\objd(#1)(#2)#3#4#5{%
  \psline[arrows={[-}, linestyle=dashed, dash=0.0 1,dashadjust=false](#1)(#2)%
  \uput{0.4}[90](#1){#3}%
  \uput{0.4}[-90](#1){#4}\uput{0.4}[-90](#2){#5}%
}

\def\eea{\end{eqnarray}}

\def\ben{\begin{displaymath}}
\def\een{\end{displaymath}}
\def\ba{\begin{array}{c}}
\def\bal{\begin{array}{l}}
\def\ea{\end{array}}

\begin{document}

\title{Finite-gap systems, tri-supersymmetry and self-isospectrality}

\author{\textsf{Francisco Correa}\textsf{$\,$, V\'{\i}t Jakubsk\'y}
\textsf{$\,$ and  Mikhail S. Plyushchay}
\\
{\small \textit{Departamento de F\'{\i}sica, Universidad de Santiago
de Chile, Casilla 307, Santiago 2,
Chile}}\\
\sl{\small{E-mails:  fco.correa.s@gmail.com, v.jakubsky@gmail.com,
mplyushc@lauca.usach.cl}}}
\date{}
\maketitle

\begin{abstract}
We show that an $n$-gap periodic quantum system with parity-even
smooth potential admits $2^n-1$ isospectral super-extensions. Each
is described by a tri-supersymmetry  that originates from a
higher-order differential operator of the Lax pair and two-term
nonsingular decompositions of it;  its local part corresponds to  a
spontaneously partially broken centrally extended nonlinear $N=4$
supersymmetry. We conjecture that any finite-gap system having
antiperiodic singlet states  admits a self-isospectral
tri-supersymmetric extension with the partner potential to be the
original one translated for a half-period. Applying the theory to a
broad class of finite-gap elliptic systems described by a
two-parametric associated Lam\'e equation, our conjecture is
supported by the explicit construction of the self-isospectral
tri-supersymmetric pairs. We find that the spontaneously broken
tri-supersymmetry of the self-isospectral periodic system is
recovered in the infinite period limit.

\end{abstract}

\section{Introduction}

Finite-gap periodic quantum systems play an important role in
physics. They underly the theory of periodic solutions in nonlinear
integrable systems, including the Korteweg-de Vries, the nonlinear
Schr\"odinger, the Kadomtsev-Petviashvili, and the sine-Gordon
equations \cite{Int, Solit, Belokolos, KriNov, Gesz}. Being
analytically solvable systems, they find various applications in
diverse areas. The list of their applications is extensive, and
among others includes the modelling of crystals \cite{Suth, bel,
AGI}, the theory of monopoles \cite{BPS}, instantons and sphalerons
\cite{Dunne, Sphalerons}, classical Ginzburg-Landau field theory
\cite{MaiSte}, Josephson junctions theory \cite{Josj}, magnetostatic
problems \cite{DobRit}, inhomogeneous cosmologies \cite{RedSh},
Kaluza- Klein theories \cite{KK}, chaos \cite{chaos}, preheating
after inflation modern theories \cite{PreHeat},  string theory
\cite{DouShe}, matrix models \cite{matrix},  supersymmetric
Yang-Mills theory \cite{SeiWit, DonWit} and AdS/CFT duality
\cite{Kazaetal}.

Some time ago it was showed by Braden and Macfarlane \cite{BraMac},
and in a more broad context by Dunne and Feinberg \cite{dunfei},
that a usual $N=2$ supersymmetric extension of a periodic quantum
system may produce a completely isospectral pair with a \emph{zero
energy doublet} of the ground states. Such a picture is completely
different from that taking place in non-periodic systems described
by the same linear $N=2$ superalgebraic structure
$\{Q_a,Q_b\}=2\delta_{ab}H$, $[Q_a,H]=0$. There, the complete
isospectrality of the super-partners happens only in the case of a
spontaneously broken supersymmetry, characterized by a positive
energy of the lowest supersymmetric doublet~\cite{Wit}. Furthermore,
it was showed that there exist peculiar isospectral supersymmetric
periodic systems,  in which the partner potentials are identical in
shape but mutually translated for a half-period, or reflected, or
translated and reflected. A pair of super-partner potentials with
such a property was named by Dunne and Feinberg as
\emph{self-isospectral}. The phenomenon of self-isospectrality with
a half-period shift was illustrated by some examples of exactly
soluble models belonging to a class of finite-gap periodic systems.
Later on isospectral and self-isospectral supersymmetric finite-gap
periodic systems were studied in various aspects \cite{noiso,
FerMiletal, sgnn}, and it was found in \cite{FerNegNie} that a
property of the self-isospectrality may also appear in some periodic
finite-gap systems based on a nonlinear supersymmetry of the second
order $\{Q_a,Q_b\}=2\delta_{ab}P_2(H)$, $[Q_a,H]=0$ \cite{nSUSY,
MPhid, KP}, where $P_2(H)$ is a quadratic polynomial. The nature and
origin of isospectrality and self-isospectrality in finite-gap
systems have remained, however, to be obscure.

Recently we showed  \cite{Selfsusy}  that  self-isospectrality may
be realized by a non-relativistic electron in the periodic magnetic
and electric fields of a special form, and indicated on a peculiar
non-linear supersymmetric structure associated with it.

In the present paper, we study superextension of quantum periodic
systems with a parity-even smooth finite-gap potential of general
form, and show that it is characterized by an unusual
tri-supersymmetric structure. This peculiar supersymmetric structure
originates from the higher order differential operator of the Lax
pair, and its decomposability in pairs of nonsingular operators. The
superalgebra, generated by three indicated integrals of motion
together with  trivial integrals associated with parity symmetry and
matrix extension, has a nonlinear nature, that reflects a
nonlinearity of a spectral polynomial of the original finite-gap
system. The higher order operator of the Lax pair of a nontrivial
$n$-gap ($n>0$) system admits $2^n-1$ nonsingular two-term
decompositions. By means of the Crum-Darboux transformation, with
each nonsingular decomposition we associate a particular
tri-supersymmetric extension, and as a result get a family of $2^n$
completely isospectral systems. We show that a local part of the
tri-supersymmetry is a spontaneously partially broken centrally
extended nonlinear $N=4$ supersymmetry, that explains the nature and
origin of the complete isospectrality. When the original finite-gap
system has in its spectrum a nonzero number of anti-periodic singlet
states corresponding to the edges of permitted bands, among all the
non-singular decompositions of the higher order operator of the Lax
pair there is a special one which corresponds to a separation of all
the singlets into orthogonal subspaces of periodic and anti-periodic
states. We conjecture that it is this separation that produces a
self-isospectral tri-supersymmetric system. This means particularly
that all the set of $2^n$ completely isospectral systems we get,
including the original $n$-gap system with the specified special
property, is divided into $2^{n-1}$ self-isospectral
tri-supersymmetric pairs. Then we apply a general theory to a broad
class of finite-gap elliptic (double periodic) quantum systems
described by a two-parametric family of associated  Lam\'e equations.
Any such a system has in its spectrum a non-empty subspace of
anti-periodic singlet states, and we support our conjecture by the
explicit construction of the self-isospectral tri-supersymmetric
pairs. We also investigate a rather intricate picture of the
infinite-period limit of the tri-supersymmetry.

The paper is organized as follows. In the next section we first
discuss general properties of the finite-gap periodic systems with a
smooth potential,  and show that any parity-even  $n$-gap system is
characterized by a hidden bosonized $N=2$ nonlinear supersymmetry of
order $2n+1$. This supersymmetry  reflects the peculiarities  of the
band structure. In Section 3 we show how the tri-supersymmetric
extensions of the system are constructed by means of the
Crum-Darboux transformation. There we also investigate a general
structure and properties of the tri-supersymmetry. In Section 4 we
apply a general theory to the case of finite-gap elliptic systems
described by the associated Lam\'e equation. In section 5, the
infinite-period limit of the tri-supersymmetry is studied. Section 6
is devoted to concluding remarks and outlook.

\section{\protect\bigskip Hidden  supersymmetry in
finite-gap systems}

To have a self-contained presentation, in this section we first
summarize briefly the properties of the quantum periodic systems
of a general form. Then we restrict the consideration to the case
of the smooth parity-even finite-gap systems to reveal in them a
hidden bosonized nonlinear supersymmetry whose order is defined by
the number of energy gaps.

\subsection{General properties of quantum periodic systems}

Consider a quantum  system given by a Hamiltonian operator
$H=-D^2+u(x)$, $D=\frac{d}{dx}$, with a  real \emph{smooth} periodic
potential $u(x)$, $u(x)=u(x+2L)$. For the corresponding stationary
Schr\"odinger equation,
\begin{equation}\label{SEH}
    H\Psi(x)=E\Psi(x),
\end{equation}
known in the literature as Hill's equation, we  choose some real
basis of solutions, $\psi_1(x;E),$ $\psi_2(x;E)$. The operator of
translation for the period $2L$, or the monodromy operator,
\begin{equation}\label{T}
    T\Psi(x)=\Psi(x+2L),
\end{equation}
commutes with the Hamiltonian $H$, $[T,H]=0$. It preserves a
two-dimensional linear vector space of solutions of (\ref{SEH}), and
can be represented there by the second order monodromy matrix
$M(E)$,
\begin{equation}\label{TM}
    T\psi_a(x;E)=\psi_a(x+2L;E)=M_{ab}(E)\psi_b(x;E).
\end{equation}
The change of the basis, $\psi_a(x;E)\rightarrow
\tilde{\psi}_a(x;E)=A_{ab}\psi_b(x;E)$, $\det A\neq 0$, generates a
conjugation of the monodromy matrix, $M(E)\rightarrow
\tilde{M}(E)=AM(E)A^{-1}$, but do not change its determinant, $\det
M(E)=\det \tilde{M}(E)$, trace, ${\rm Tr} M(E)={\rm Tr}
\tilde{M}\equiv \mathcal{D}(E)$, and eigenvalues, given by solutions
of the characteristic equation
\begin{equation}\label{chareq}
    \det (M(E)-\mu I)=0.
\end{equation}
Let us choose a particular basis of solutions fixed by  conditions
\begin{equation}\label{psi0}
    \psi_1(0;E)=1,\quad \psi'_1(0;E)=0,\qquad
    \psi_2(0;E)=0,\quad \psi'_2(0;E)=1,
\end{equation}
where prime denotes the $x$-derivative. Differentiating  relation
(\ref{TM}) in $x$ and putting then $x=0$ in (\ref{TM}) and in the
derived relation, we find the form of the monodromy matrix in basis
(\ref{psi0}),
\begin{equation}\label{monod}
    M(E)=\left(%
    \begin{array}{cc}
 \psi_1(2L; E) &  \psi'_1(2L;E) \\
    \psi_2(2L;E) &  \psi'_2(2L;E) \\
    \end{array}%
    \right).
\end{equation}
Wronskian $W(\psi_1,\psi_2)=\psi_1\psi'_2-\psi'_1\psi_2$  of any two
linearly independent solutions of equation (\ref{SEH}) takes a
nonzero $x$-independent value, which for basis (\ref{psi0}) is equal
to $1$. Then the explicit form of the real monodromy matrix
(\ref{monod}) shows that a basis-independent value of its
determinant does not depend on energy either, $\det M(E)=1$, and so,
$M(E)\in sl(2,\R)$. Note that the change $x=0\rightarrow x_0\in\R$
in  relations (\ref{psi0}) gives a one-parametric family of the
bases, $\psi_a(x;x_0,E)=A(x_0)_{ab}\psi_b(x;E)$, $A(x_0)\in
sl(2,\R)$, playing an important role in the theory of periodic
quantum systems \cite{Int}-\cite{Belokolos}. In such a basis, the
monodromy matrix will include an additional dependence on the marked
point $x_0$, $M(E,x_0)\in sl(2,\R)$.

With taking into account that $\det M=1$, the characteristic
equation (\ref{chareq}) is reduced to $1-\mathcal{D}(E)\mu+\mu^2=0$,
and the basis-independent eigenvalues of the monodromy matrix are
given in terms of its trace\footnote{The trace of the monodromy
matrix is called in the literature the Lyapunov function, Hill
determinant, or discriminant of the Schr\"{o}dinger equation.},
\begin{equation}
    \mu_{1,2}(E)=\frac{1}{2}\mathcal{D}(E)\pm
    \sqrt{{\mathcal{D}(E)}^2/4-1}.
    \label{disc}
\end{equation}
In correspondence with $\det M(E)=1$,  $\mu_1\mu_2=1$.
Common
eigenstates of $H$ and $T$ are described by the Bloch-Floquet
functions $\psi_{\pm}(x;E)$, which satisfy a relation
\begin{equation}\label{kappa}
    T\psi_{\pm}(x;E)=\exp(\pm i
    \kappa(E))\psi_{\pm}(x;E),
\end{equation}
where $\mu_{1,2}(E)=\exp(\pm i \kappa(E))$, and the quasi-momentum
$\kappa(E)$ is given by
\begin{equation}\label{kapE}
   2\cos \kappa(E)= \mathcal{D}(E).
\end{equation}

\begin{figure}[h!]
\begin{center}
\includegraphics[scale=0.8]{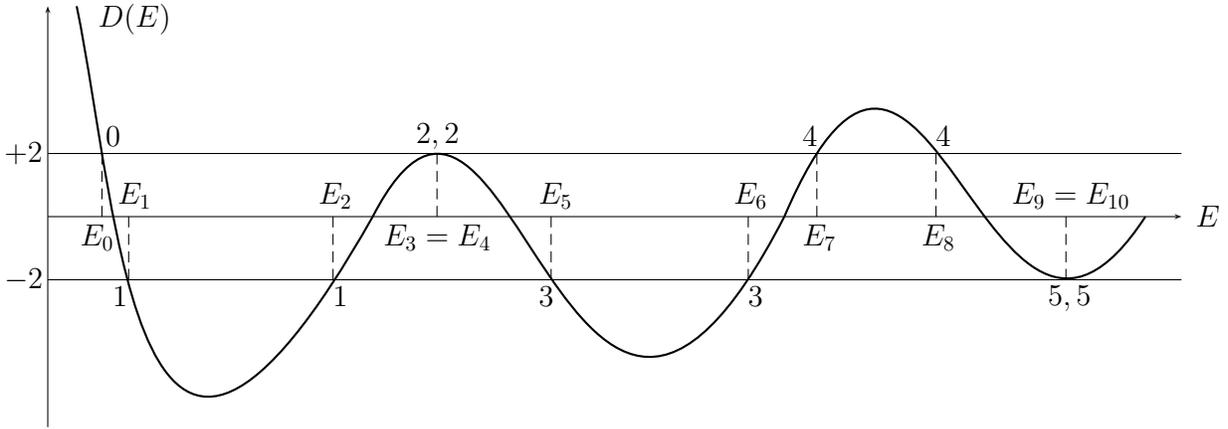}
\caption{The discriminant $\mathcal{D}(E)$ in a generic situation
of a periodic potential.} \label{dis}
\end{center}
\end{figure}

The values of the discriminant $\mathcal{D}(E)$  define the spectral
properties of the periodic Schr\"{o}dinger equation. For some
energies $E\in(E_{2i-1},E_{2i})$, $i=0,\ldots$, $E_{i}<E_{i+1}$,
$E_{-1}=-\infty$, the quasi-momentum $\kappa(E)$ takes complex
values, and $|\mathcal{D}(E)|>2$. Solutions corresponding to such
$E$'s are not physically acceptable as they diverge in $x=-\infty$
or $+\infty$. For these values of $E$ we have a forbidden band, or
an energy gap, see Fig. \ref{dis}. In a generic case, a periodic
quantum system  has an infinite number of gaps. The width of the
gaps decreases rapidly when energy increases, while the rate of
decrease depends on the smoothness of the potential. In the case of
analytic potentials, the gaps decrease exponentially. Energies $E$
for which $|\mathcal{D}(E)|\leq 2$, define permitted bands, or
permitted zones. Here, the quasi-momentum $\kappa(E)$ takes real
values, and complex numbers $\exp(\pm i \kappa(E))$ have modulus
equals to 1. All the energy levels with $|\mathcal{D}(E)|<2$ are
doubly degenerate, but  for $|\mathcal{D}(E)|=2$ we have two
essentially different cases. For those $E$, which separate permitted
and prohibited bands, corresponding eigenvalue of the monodromy
matrix is \emph{non-degenerate}, the matrix $M$ has a form of Jordan
matrix, and a physical singlet band-edge state is periodic, $\exp(i
\kappa(E))= +1$, if $\mathcal{D}(E)=2$, while for
$\mathcal{D}(E)=-2$ a singlet state is antiperiodic, $\exp(i
\kappa(E))= -1$. When $|\mathcal{D}(E)|=2$ but the corresponding
eigenvalue of the monodromy matrix is doubly degenerate, $M$ is
diagonalizable  on the two linearly independent Bloch-Floquet
states, which both are periodic if $\mathcal{D}(E)=2$, or are
antiperiodic when $\mathcal{D}(E)=-2$. This second situation, that
corresponds to points $E_3=E_4$ and $E_9=E_{10}$ on Fig. \ref{dis},
takes place when a prohibited band disappears.

Summarizing, the interval $(-\infty,E_0)$ constitutes the lowest
forbidden band. The permitted bands with $|\mathcal{D}(E)|\leq 2$
are separated by prohibited bands, or energy gaps. All the energy
levels in the interior of permitted bands have a double
degeneration, while the states at their edges  are singlets.

According to the oscillation theorem \cite{Lameas}, the common
eingenstates of $H$ and monodromy operator $T$ with energies
$E_0  <   E_1\leq E_2  <   E_3\leq E_4  <   E_5\leq E_6 <  \ldots $ such
that $|\mathcal{D}(E_k)|=2$, are described by the wave functions
which are characterized by the  periods
$2L,\,4L,\,4L,\,2L,\,2L,\,4L,\,4L...$, and by the node numbers in
the period $2L$ equal to $0,1,1,2,2,3,3,\ldots$, see Fig. \ref{dis}.
The odd number of nodes corresponds to antiperiodic states, whereas
the periodic states have an even number of nodes in the period $2L$.
The singlet states at the edges of the same prohibited band have the
same number of nodes and the same periodicity, and their nodes are
alternating.

\subsection{Finite-gap systems and hidden bosonized supersymmetry}

In some periodic potentials infinite number of bands merge together
so that only finite number of gaps remains in the spectrum. Such
potentials are called finite-gap. The simplest case of a zero-gap
system corresponds here to a free particle with
$u(x)=const$~\footnote{Here and in what follows we do not count the
prohibited band $(-\infty,E_0)$ that always presents in any periodic
system with a smooth potential.}. For the Schr\"{o}dinger equation
with a finite-gap potential the spectrum and eigenfunctions can be
presented in an analytical form~\footnote{In this sense, and in a
contrast with, for example, the Kronig-Penney model, finite-gap
potentials play the same role in solid-state physics as the Kepler
problem in atomic theory.}. Having also in mind that for analytical
potentials the size of the gaps decreases exponentially when energy
increases, any periodic potential can be approximated by a
finite-gap potential if narrow gaps are disregarded.

From now on we suppose that a periodic potential $u(x)$ is
finite-gap. \emph{Additionally}, we  assume  that it is an
\emph{even} function, $u(x)=u(-x)$. Then a reflection (parity)
operator $R$, $R\psi(x)=\psi(-x)$, is a nonlocal  integral of
motion, $[R,H]=0$. Periodicity and parity symmetry together imply
that the potential possesses also a middle-point reflection symmetry
$u(L+x)=u(L-x)$.

The spectrum $\sigma(H)$ of a nontrivial  $n$-gap ($n>0$) system is
characterized by the band structure,
$\sigma(H)=[E_{0},E_{1}]\cup\ldots\cup
[E_{2n-2},E_{2n-1}]\cup[E_{2n},\infty)$, where $E_0<E_1<\ldots
<E_{2n}$ are the non-degenerate energies corresponding to the $2n+1$
\emph{singlet} band-edge states $\Psi_i(x)$, $H\Psi_i=E_i\Psi_i$,
$i=0,1,\ldots,2n$. Since parity operator $R$ is an integral, each
singlet state $\Psi_i(x)$ has a definite parity, $+1$ or $-1$. The
energy levels in the interior of permitted bands, $E\in
(E_{2i},E_{2i+1})$, $i=0,\ldots, n$, are doubly degenerate, and
certain linear combinations of corresponding Bloch-Floquet doublet
states are the eigenstates of $R$ with eigenvalues $+1$ and $-1$.
These properties  indicate on the presence of a \emph{hidden
bosonized $N=2$ supersymmetry} in any finite-gap system, for which
operator $R$ has to play  the role of the grading operator.  The
presence of $2n+1\geq 3$ singlet states indicates, however, on its
nonlinear nature \cite{nSUSY,MPhid,KP}. The supercharges and the
form of the corresponding nonlinear superalgebra can easily be
identified.

Any finite-gap system is characterized by the presence of a nontrivial
integral of motion in the form of an anti-Hermitian
differential operator of order $2n+1$,
\begin{equation} \label{prez}
    A_{2n+1}=D^{2n+1}+c_{2}^A(x)D^{2n-1}+c_{3}^A(x)D^{2n-2}+\ldots
    +c_{2n}^A(x),
\end{equation}
where the coefficient functions $c_i^A(x)$ are real. The absence
of the term proportional to $D^{2n}$ in its structure, i.e. an
equality $c_{1}^A(x)=0$, is dictated by the condition
$[A_{2n+1},H]=0$. Other coefficients $c_j^A(x)$ are fixed in the
form of  polynomials in the potential $u(x)$ and its derivatives
\cite{Gesz}. Thus, for periodic potential, $A_{2n+1}$ is a
periodic operator, i.e. $[A_{2n+1},T]=0$. $( A_{2n+1},H)$ is known
as the Lax pair of the $n$-th order Korteweg-de Vries (KdV)
equation. A possible form of the $n$-gap potential is fixed  by a
nonlinear equation, which has a sense of the $n$-th equation of
the stationary KdV hierarchy \cite{Solit,Gesz}. This equation can
be represented alternatively as
\begin{equation}\label{n-gap}
    \tilde{L}(J\tilde{L})^n 1=0,\quad
    \tilde{L}=D^3+2(uD+Du),
\end{equation}
where $J$ is the operator of indefinite integration $J=D^{-1}$
\cite{Belokolos}. The form of a \emph{one-gap} potential is fixed by
this equation in a unique manner,
\textbf{$u(x)=2\mathcal{P}(x+\omega_2 +c)$}, where $\mathcal{P}(x)$
is the doubly periodic (elliptic) Weierstrass function \cite{ww},
and  $c$ is a  constant. To have a real-valued potential, one of the
periods of $\mathcal{P}(x)$ is chosen to be real, $2\omega_1=2L$,
while another period $2\omega_2$ is assumed to be pure imaginary,
and $c\in \R$. In the case $n>1$ the form of the potential $u(x)$ is
not fixed uniquely even if it is restricted to a class of elliptic
functions.

The mutually commuting operators $A_{2n+1}$ and $H$ satisfy the
relation
\begin{equation} \label{spectral}
    -A_{2n+1}^2=P_{2n+1}(H),\qquad P_{2n+1}(H)=\prod_{j=0}^{2n}(H-E_j),
\end{equation}
where $P_{2n+1}(H)$ is a spectral polynomial given in terms of
singlet energies. It is in accordance with Burchnall-Chaundy
theorem \cite{BU-CH, Ince}, which says that if two differential in
$x$  operators $A$ and $B$ of mutually prime orders $l$ and $k$,
respectively,  commute, $[A,B]=0$, they satisfy a relation
$P(A,B)=0$, where  $P$  is a polynomial  of order $k$ in $A$, and
of order $l$ in $B$.  Equation (\ref{spectral}) corresponds to a
non-degenerate ($E_i\neq E_j$ for $i\neq j$) spectral elliptic
curve of genus $n$  associated with an $n$-gap periodic system
\cite{Int}-\cite{Belokolos}~\footnote{Because of the described
properties, $u(x)$ is called algebro-geometric finite-gap
potential.}.

As a consequence of (\ref{spectral}), the operator $A_{2n+1}$
annihilates  all the  $2n+1$ singlet band-edge states. Indeed, from
$[A_{2n+1},H]=0$ we have $A_{2n+1}\Psi_j=\alpha\Psi_j+\beta\Phi_j$,
where $\Psi_j$
 is a physical ($T\Psi_{j}=\gamma \Psi_{j},\ \gamma\in\{-1,1\}$)
 and $\Phi_j$ is a non-physical solution
 corresponding to a band-edge energy $E_j$.
Acting from the left by $T$, we get  $\gamma A_{2n+1}\Psi_j=
\gamma\alpha\Psi_j+\beta T\Phi_j$, and, therefore,  $\beta (\gamma
T-1)\Phi_j=0.$ As $\Phi_j$ is neither periodic nor antiperiodic,
the last equation can be satisfied if and only if $\beta=0$. Then,
equation (\ref{spectral}) dictates that $\alpha=0$.

Consider the Wronskian of the singlet states, $W^A\equiv
W(\Psi_0,...,\Psi_{2n})$. In a generic case the Wronskian of $s$
linearly independent  functions  that form a kernel of an
arbitrary linear differential operator of order $s$, ${\cal
L}=D^s+c_1(x)D^{s-1}+\ldots$,  satisfies the Abel identity
$W'(x)=-c_1(x)W$ \cite{Ince}. For operator (\ref{prez}) a
corresponding coefficient function is $c^A_1(x)=0$, and because of
the linear independence of the singlet band-edge states we find
that
\begin{equation}\label{WronC}
    W^A(x)=C\neq 0,
\end{equation}
where $C$ is a constant.
 When $s$ linearly independent zero modes $\varphi_j$, $j=1,\ldots,s$,
of operator ${\cal L}$  are known, the form of this operator  can be
reconstructed in their terms. The coefficients $c_{k}(x)$ are
defined by relations
 $   c_{k}(x)=-\frac{W_{k}}{W},\qquad
     k=1,\ldots,s,$
where the functions $W_{k}(x)$ are obtained from Wronskian
$W=W(\varphi_1,\ldots,\varphi_s)$ by replacing in it $\varphi
_{j}^{(s-k)}\equiv D^{s-k}\varphi_j$ by $\varphi _{j}^{(s)}$
\cite{DarSol} , see Appendix A. In our case, each singlet band-edge
state $\Psi_i(x)$, being a zero mode of $A_{2n+1}$, possesses a
definite parity. As a result, with taking into account
(\ref{WronC}),  we find that the coefficients $c_{2r}^{A}(x)$ are
\emph{odd}, while the coefficients $c_{2r+1}^{A}(x)$ are \emph{even}
non-singular functions. Hence the integral $A_{2n+1}$ is parity odd,
\begin{equation}\label{amonRA}
    \{R,A_{2n+1}\}=0.
\end{equation}
 Introducing two Hermitian operators

\begin{equation}\label{ZAR}
    Z=Z_1=iA_{2n+1}, \quad Z_2=iRZ,
\end{equation}
 and identifying them as odd supercharges,
 we conclude that any finite-gap periodic system with even smooth potential is characterized by
 a hidden bosonized nonlinear $N=2$ supersymmetry
of order $2n+1$ \cite{MPhid,KP,BosoSusy},
\begin{equation}\label{ZZH}
    \{Z_a,Z_b\}=2\delta_{ab}P_{2n+1}(H) , \quad a,b=1,2.
\end{equation}

\section{Tri-supersymmetric extensions of finite-gap systems}
In this section we show that the application of a non-singular
Crum-Darboux transformation to a finite-gap system produces a
partner system with identical spectrum, and study a peculiar
supersymmetry appearing in the obtained isospectral pair.

\subsection{Darboux transformations and supersymmetry}

A usual model of supersymmetric quantum mechanics is based on a
Darboux  transformation, by which an (almost) isospectral system can
be associated with a given quantum system.

Consider a Hamiltonian $H=-\frac{d^2}{dx^2}+u(x)$, and an eigenstate
$\psi_\star$ corresponding to  a fixed eigenvalue $E_\star$,
$H\psi_\star=E_\star\psi_\star$. Here we do not assume any
regularity conditions for  $\psi_\star$. It can be a physical
eigenstate, or a second, non-physical solution of the second order
differential equation, corresponding to a physical energy level
$E_\star$, or can be a solution corresponding to a nonphysical value
$E_\star$. The Darboux transformation is generated by a first order
differential operator $A_1=\frac{d}{dx}-{(\ln
\psi_\star)}^{\prime}$, which annihilates $\psi_\star$,
$A_1\psi_\star=0$, and relates $H$ with another Hamiltonian
\begin{equation}
    \tilde{H}=-\frac{d^2}{dx^2}+\tilde{u}(x),
    \qquad \tilde{u}(x)=u(x)-2\frac{d^2}{dx^2}\ln \psi_\star,
    \label{darboux1}
\end{equation}
 by means of an intertwining relation
\begin{equation}\label{darboux2}
    A_1H= \tilde{H} A_1.
\end{equation}
Then for eigenstates of two Hamiltonians corresponding to the same
arbitrary value of energy $E$, we have
\begin{equation}\label{DA1}
    H\psi_E=E\psi_E, \qquad \tilde{H} \tilde \psi_E=E \tilde \psi_E,
\end{equation}
\begin{equation}\label{DA2}
    \tilde \psi_E=\frac{1}{\sqrt{E-E_\star}} A_1 \psi_E,
    \qquad \psi_E= \frac{1}{\sqrt{E-E_\star}}A^{\dagger}_1
     \tilde \psi_E\, .
\end{equation}
Relations (\ref{DA1}), (\ref{DA2})  have a symmetry
$H\leftrightarrow \tilde{H}$, $\psi_E\leftrightarrow
\tilde{\psi}_E$, $A\leftrightarrow A^\dagger$. This  reflects a
property that the transformation corresponding  to the adjoint
intertwining relation
\begin{equation}\label{invD}
    A_1^\dagger\tilde{H}=HA_1^\dagger
\end{equation}
is generated by the operator $A^\dagger_1$, which annihilates a
state $\tilde{\psi}_\star=1/{\psi_\star}$,
$A^\dagger_1\tilde{\psi}_\star=0$, and acts in an opposite direction
by relating $\tilde{H}$ with $H$. It is easy to see that the both
Hamiltonians can be represented in terms of operators $A_1$ and
$A_1^\dagger$,
 $   H=A^\dagger_1 A_1 +E_\star,\qquad \tilde{H}=A_1A_1^\dagger
    +E_\star.$

 Usually, the Darboux transformation is chosen to
annihilate a nodeless physical ground state $\psi_0$ with energy
$E_0$. In such a case, the potentials $u(x)$ and $\tilde{u}(x)$ are
both smooth and regular, or both have the same
singularities~\footnote{Singular Darboux transformations generated
by the states with nodes also find some applications,
see~\cite{CKS}.}. In a \emph{non-periodic} case, the physical
nodeless ground state $\psi_0$ vanishes at the ends of a (possibly
infinite) interval. As a consequence, there is no physical partner
state with the same energy in the spectrum of $\tilde{H}$. Indeed,
the state $\tilde{\psi}_0=1/\psi_0$ annihilated by $A^\dagger_1$ is
divergent at infinity and is not physical. In this case both systems
are almost isospectral, their spectrum is the same except the energy
level $E_0$ to be absent from the spectrum of $\tilde{H}$. Note that
from the viewpoint of the adjoint intertwining relation
(\ref{invD}), the transformation from $\tilde{H}$ to $H$ is
generated by the operator $A_1^\dagger$ associated with a
nonphysical state $\tilde{\psi}_0=1/\psi_0$, which corresponds to a
nonphysical for $\tilde{H}$ eigenvalue $E_0$. On the other hand, in
correspondence with (\ref{DA2}), for $E=E_0$ we still have relations
$\psi_0=A^\dagger_1\tilde{\eta}_0$ and
$\tilde{\psi}_0=1/{\psi_0}=A_1\eta_0$, but
 $   \tilde{\eta}_0=-\frac{1}{\psi_0} \int^x\psi_0^2(x)dx,\quad
    \eta_0=\psi_0\int^x \psi_0^{-2} dx$
are the non-physical, non-normalizable solutions of the equations
$H\eta_0=E_0\eta_0$ and $\tilde{H}\tilde{\eta}_0=E_0\tilde{\eta}_0$.

In the periodic case with $\psi_0$ corresponding to the singlet
band-edge state of the lowest energy, $\eta_0$ and $\tilde{\eta}_0$
are the non-physical, non-periodic divergent solutions. {}From this
discussion it is also clear that if the Darboux transformation is
realized with a nodeless state $\psi_\star$ such that both states
$\psi_\star$ and $1/\psi_\star$ are not physical (non-normalizable),
the energy level $E_\star$ is absent from  the spectra of both
partner systems, and physical energy levels satisfy a relation
$E>E_\star$.

The relation between the Darboux transformation and the usual
supersymmetric quantum mechanics is direct.  The Hamiltonians $H$
and $\tilde{H}$ shifted for the constant $E_\star$ are known as
superpartner Hamiltonians, and form a superextended system described
by the matrix Hamiltonian
\begin{equation}\label{Hsusy}
    H=\left(
    \begin{array}{cc}
      H_+ & 0 \\
      0 & H_- \\
    \end{array}
    \right),
\end{equation}
where
\begin{equation}\label{HHEE}
    H_-\equiv H-E_\star=-\frac{d^2}{dx^2}+W^2-W',\quad H_+\equiv
    \tilde{H}-E_\star=-\frac{d^2}{dx^2}+W^2+W',
\end{equation}
 and  $W(x)$ is a
superpotential, $W=-\frac{d}{dx}\ln\psi_\star$.

With the Darboux transformation, two Hermitian linear differential
matrix operators
\begin{equation}\label{Q12susy}
Q_1=\left(
\begin{array}{cc}
  0 & A_1 \\
  A^{\dagger}_1 & 0 \\
\end{array}
\right), \qquad Q_2=i\sigma_3Q_1
\end{equation}
are associated,  in terms of which intertwining relations
(\ref{darboux2}) and (\ref{invD}) take a form of conservation laws
for supercharges $Q_a$, $[Q_a,H]=0$, $a=1,2$. Together with
Hamiltonian (\ref{Hsusy}) they generate the linear $N=2$
superalgebra
\begin{equation}\label{linsusy}
[Q_a,H]=0,\quad \{Q_a,Q_b\}=2\delta_{ab}H.
\end{equation}
A diagonal Pauli matrix $\sigma_3$ plays here the role of the
grading operator, $[\sigma_3,H]=0$, $\{\sigma_3,Q_a\}=0$.

In a non-periodic case, if $\psi_\star$ or $1/\psi_\star$ is
normalizable, there exists a two-component physical state
annihilated by both matrix supercharges, which is a ground state of
zero energy of one of the super-partner subsystems. It is invariant
under corresponding supersymmetry transformations generated by
$Q_a$, and we have the case of exact, unbroken supersymmetry. The
supersymmetric doublets of states corresponding to positive energies
are mutually transformed by supercharges $Q_a$ in correspondence
with (\ref{DA2}). In the case if both $\psi_\star$ and
$1/\psi_\star$ are not physical, all the states of supersymmetric
system (\ref{Hsusy}) are organized in supersymmetric doublets,
including the states of the lowest energy, that takes here a
nonzero,  positive value. This picture corresponds to the broken
supersymmetry, which describes a pair of completely isospectral
super-partner systems.

In the case of a periodic quantum system with a smooth potential,
the supersymmetric system (\ref{Hsusy}) constructed on the base of
the Darboux transformation with a nodeless singlet band-edge state
$\Psi_0$ will be characterized by zero energy doublet of the states
given by the columns  $(0,\Psi_0)^t$ and $(1/\Psi_0,0)^t$. Both
these states are annihilated by the supercharges $Q_a$, and
corresponding $N=2$ supersymmetry is unbroken. Here the
super-partner systems are completely isospectral as in the
non-periodic case with broken supersymmetry.

\subsection{Higher-order Crum-Darboux transformations}

(Almost) isospectral systems can also  be related by differential
operators of higher order, that corresponds to the situation well
described by the generalization of the Darboux transformation due to
Crum.

Let a differential  operator $A_k$ of order $k$,
$A_k=D^{k}+\sum_{j=1}^kc^A_jD^{k-j}$,  annihilates a space $V$
spanned by $k$  eigenstates of the Hamiltonian $H$, $V=span\,
\{\psi_1,\ldots,\psi_k\}$, which are not obligatory to be physical. Then, there holds the relation
\begin{equation}\label{crum}
    A_kH=\tilde{H}A_k,\quad  \tilde{H}=H+2(c^A_1)'=H -2(\ln
    W(\psi_1,\ldots,\psi_k))''.
\end{equation}
In the case of the Darboux transformation ($k=1$), the Wronskian of
a single function is the function itself, and   (\ref{crum}) reduces
to the intertwining relation of the standard supersymmetry. The
operators $A_k$ and $A_k^{\dagger}$ produce the relations of the
form (\ref{DA2}) for energies $E\ne E_i$, $i=1,\ldots,k$. For $k>1$,
Eq. (\ref{crum}) underlies a higher-order (nonlinear) generalization
of supersymmetric quantum mechanics, see \cite{nSUSY,MPhid,KP}. In a
generic case the spectra of $H$ and $\tilde H$ are almost identical,
their spectra can be different in $k$ or less number of physical
eigenvalues. For a quantum system described by $H$, one can obtain
various partner Hamiltonians $\tilde{H}$, by choosing different sets
of eigenstates $\psi_1,\ldots,\psi_k$. However, if we want to get
the associated partner Hamiltonian $\tilde{H}$ with the same
regularity properties as the initial Hamiltonian $H$, these states
have to be chosen in a special way.

The higher-order Crum-Darboux transformations can be factorized
into the consecutive chain of the first order Darboux, or the
second-order Crum transformations, see  \cite{irre}. We are
interested in the conditions which not only ensure the regularity
of the transformations for a periodic even finite-gap system,  but
also produce an isospectral  \emph{even} partner potential. In a
periodic system, a regular transformation of the second order can
be obtained if the kernel of the operator $A_2$ consists of the
states corresponding to the edges of the same prohibited band
\cite{samsonov}. These two states, due to the oscillation theorem
mentioned in the previous section, have the same period and the
same number of alternating nodes. The Wronskian of the functions
selected in this way is a function of a definite sign, not taking
the zero value. Consider a Crum-Darboux transformation that
annihilates the indicated pairs of the edge-states.  In the case
if the order of the transformation is odd, it has also to
annihilate the nodeless ground state $\Psi_0$. This guarantees a
smooth and singularity-free potential of the partner Hamiltonian.

Concluding, we can construct a whole family of the partner
finite-gap periodic systems  by means of the Crum-Darboux
transformations, just by choosing appropriately the singlet states
of the original system, respecting the rules described above. For
instance, the generator of a hidden bosonized supersymmetry  $Z$
produces a Crum-Darboux transformation associated with the
\emph{trivial} selection: it annihilates all the singlets. Since the
Wronskian of all the singlet states is a nontrivial constant, see
Eq. (\ref{WronC}), we find that the partner Hamiltonian coincides
with the original one,  and the intertwining relation reduces to the
relation of commutation of $Z$ and $H$.

Below,  by means of  nontrivial Crum-Darboux transformations, we
shall construct a family of $2^n-1$ different completely isospectral
partner systems for a given arbitrary $n$-gap periodic system, and
reveal a special nonlinear supersymmetry appearing in any pair of
the total family of $2^n$ systems. The key role in the construction
will belong to the already described hidden bosonized supersymmetry.

It is worth to notice here that a regular Crum-Darboux
transformation for finite-gap periodic systems can also be
produced by making use of certain Bloch functions
\cite{FerMiletal,sgnn,FMRS,FerGan1, FerGan2}. A partner for
parity-even potential obtained in such a way in a generic case,
however, will not be an even function. We shall return to this
point in the last section.

\subsection{Tri-supersymmetric extensions and centrally
extended nonlinear $N=4$ supersymmetry}

Consider an $n$-gap periodic system, and mark $r\leq n$ prohibited
bands in its spectrum. $2r$ singlet physical states at the edges of
these prohibited bands span a $2r$-dimensional linear vector space
which we denote by $V_{+}$. The Wronskian of the corresponding $2r$
singlet band-edge states is a  nodeless $2L$-periodic even function.
Let ${Q_+}$ be a linear differential operator of order $2r$ that
annihilates the space $V_+$,
 \begin{equation}\label{Q+}
    {Q_+}=D^{2r}+\sum_{j=1}^{2r}c_j^{+}(x)D^{2r-j},\quad
    {Q_+}V_{+}=0.
\end{equation}
Singlet band-edge states are periodic or anti-periodic, and can be
presented by real wave functions. So, the coefficient functions in
(\ref{Q+}) are real. Taking also into account that any band-edge
state has a definite parity, one can show that (\ref{Q+})  is a
$2L$-periodic  even  operator, $[T,Q_+]=[R,Q_+]=0$, see Appendix A.

The kernel of the integral $Z$ has a form $\mbox{Ker}\, Z =V_+\oplus
V_-$, where $V_-$ is a supplementary $2(n-r)+1$-dimensional linear
vector space spanned by the rest of singlet band-edge states. Then
$Z$ can be decomposed as $Z={S}^{\dagger}{Q_+}$, where $S^\dagger$
is a differential operator of order $2(n-r)+1$ with the property
$S^\dagger Q_+ V_-=0$. Hermiticity of $Z$ and Eqs.
(\ref{prez}), (\ref{ZAR}), (\ref{Q+}) mean that
 \begin{equation}\label{zab}
    Z={S}^{\dagger}{Q_+}=Q_+^{\dagger}{S},
\end{equation}
and
\begin{equation}\label{Bcj}
    -iS=D^{2(n-r)+1}+\sum_{j=1}^{2(n-r)+1}c_j^{S}(x)D^{2(n-r)+1-j},
\end{equation}
where the coefficient functions are real and $c^S_1(x)=c_1^+(x)$.
{}From the properties of $Z$ and $Q_+$, we also find that $S$ is a
$2L$-periodic parity-odd operator.

Now we show that $\mbox{Ker}\, S=V_-$. To this end we note that in
accordance with Eq. (\ref{crum}) and the Abel identity
$W'=-c_1(x)W$, the equality $c_1^S(x)=c_1^+(x)$ obtained directly from (\ref{zab}) means that the
application of the Crum-Darboux transformations with the operators
$Q_+$ and $S$ produces the same non-singular super-partner
Hamiltonian $\tilde{H}=H+2(c_1^+)'$ satisfying the intertwining
relations
\begin{equation}\label{inter1}
     {Q_+}H=\tilde{H}{Q_+},\qquad {S}H=\tilde{H}{S},
\end{equation}
\begin{equation}\label{inter2}
     Q_+^{\dagger}\tilde{H}=HQ_+^{\dagger},\qquad
     {S}^{\dagger}\tilde{H}=H{S}^{\dagger}.
\end{equation}
The Hamiltonian $\tilde{H}$ describes a periodic system with an even
potential of the period $2L$ with $n$ gaps in the spectrum. Thus,
there exists an odd Hermitian differential operator $\tilde{Z}$ of
the form (\ref{prez}) commuting  with $\tilde{H}$. The intertwining
relations (\ref{inter1}), (\ref{inter2}) provide an alternative,
two-term decomposed form for $\tilde{Z}$. Indeed, we get
$[\tilde{H},S Q^{\dagger}_+]=[\tilde{H},Q_+S^{\dagger}]=0$. Both
operators $SQ^{\dagger}_+$ and $Q_+S^{\dagger}$ are of order $2n+1$,
and should coincide with   $\tilde{Z}$ up to some polynomial in
$\tilde{H}$. However, they anticommute with reflection operator $R$,
and this implies that
 \begin{equation}\label{zzab}
     \tilde{Z}=SQ^{\dagger}_+=Q_+S^{\dagger}.
\end{equation}
Take any of $2(n-r)+1$ singlet states $\Psi_-\in V_-$  not
annihilated by $Q_+$.  Multiplying the equation $Z\Psi_-=0$ by $Q_+$
from the left and using (\ref{zab}) and (\ref{zzab}), we get
$SQ_+^{\dagger}Q_+\Psi_-=0$. The operator $Q^{\dagger}_+Q_+$ is
$2L$-periodic operator,  and as it follows from (\ref{inter1}),
(\ref{inter2}), commutes with the Hamiltonian $H$. It changes
neither energy nor the period of a singlet state $\Psi_-$, and then
$Q^{\dagger}_+Q_+\Psi_-=\alpha\Psi_-$, where $\alpha$ is a non-zero
number. Hence,
$
    SQ_+^{\dagger}Q_+\Psi_-=\alpha S \Psi_-=0,
$ and we conclude that $\mbox{Ker}\, S=V_-$. Changing the notation,
$Q_-\equiv S$, we have
\begin{equation}\label{ZZ}
    {Z}=Q_+^{\dagger}Q_-=Q_-^{\dagger}Q_+,\qquad
    \tilde{Z}=Q_+Q_-^{\dagger}=Q_-Q_+^{\dagger},
\end{equation}
and
\begin{equation}
    \mbox{Ker}\ Q_+\oplus \mbox{Ker}\ Q_-=\mbox{Ker}\
    Z,\qquad \mbox{Ker}\ Q_+^\dagger \oplus \mbox{Ker}\ Q_-^\dagger=\mbox{Ker}\
    \tilde{Z}.
\end{equation}
This result means a complete isospectrality of the finite-gap
periodic systems described by the Hamiltonians $H$ and $\tilde{H}$.
Indeed, in accordance with the properties of the Crum-Darboux
transformations, the action of both operators $Q_+$ and $Q_-$ on any
doublet of eigenstates of $H$ from the interior of permitted bands
transforms it  into a doublet of eigenstates of $\tilde{H}$ with the
same energy value. The adjoint operators $Q_+^\dagger$ and
$Q_-^\dagger$ act in the same way in the opposite direction. The
singlet states of $H$ annihilated by $Q_+$ (or $Q_-$) are
transformed by $Q_-$ (or $Q_+$) into zero modes of $Q_+^\dagger$ (or
$Q_-^\dagger$) being the singlet states of $\tilde{H}$ of the same
energy. The same picture is valid for the singlet states of
$\tilde{H}$ annihilated by $Q_+^\dagger$ (or $Q_-^\dagger$) and
transformed by $Q_-$ (or $Q_+$) into the corresponding singlet
states of $H$.

The intertwining relations (\ref{inter1}), (\ref{inter2})  as well
as factorization of the supercharges of bosonized supersymmetry
can be rewritten in a compact form once we use the matrix
formalism and define an extended Hamiltonain $\mathcal{H}$ and
operators $\mathcal{Q}_{\pm}$ and $\mathcal{Z}$,
\begin{equation}
    \mathcal{H}=\left(
    \begin{array}{cc}
    \tilde{H} & 0 \\
    0 & H%
    \end{array}%
    \right), \quad  \mathcal{Q}_{\pm}=\left(
    \begin{array}{cc}
    0 & Q_{\pm} \\
    Q_{\pm}^{\dagger } & 0%
    \end{array}%
    \right),\quad \mathcal{Z}=\left(
    \begin{array}{cc}
    \tilde{Z} & 0 \\
    0 & Z%
    \end{array}%
    \right)\label{matHQZ}.\end{equation}
Here relations (\ref{inter1}), (\ref{inter2}) and (\ref{ZZ}) can be
presented as
\begin{equation}\label{ZQH}
  \left[\mathcal{H}, \mathcal{Z}\right] =0,\quad
  \left[ \mathcal{H},\mathcal{Q}_{\pm}\right]=0,
\end{equation}
\begin{equation}\label{ZQQ}
   \mathcal{Z}=
  \mathcal{Q}_{-}\mathcal{Q}_{+}=\mathcal{Q}_{+}\mathcal{Q}_{-}.
\end{equation}
The triplet  $\mathcal{Q}_{+}$, $\mathcal{Q}_{-}$ and $\mathcal{Z}$
is a set of \emph{commuting} integrals for the superextended system
described by the matrix  Hamiltonian $\mathcal{H}$. There exists a
common basis, in which $\mathcal{Q}_{\pm}$, $\mathcal{Z}$ and
$\mathcal{H}$ are diagonal, and since all these operators are
self-adjoint, their eigenvalues are real. We have a chain of
equalities
\begin{equation}
     \mathcal{Z}^2=\mathcal{Q}_+\mathcal{Q}_-\mathcal{Q}_+\mathcal{Q}_-
     =\mathcal{Q}^2_+\mathcal{Q}^2_-=
     P_{\mathcal{Z}}(\mathcal{H})=\prod_{j=1}^{2n+1}(\mathcal{H}-E_j),
\label{Z2}
\end{equation}
where $P_{\mathcal{Z}}$ is a positive-semidefinite spectral
polynomial, and $E_j$ are energies of the singlet states of
subsystems. In correspondence with (\ref{Z2}), the band-edge states
of the extended system  are organized in supersymmetric doublets, on
which $\mathcal{Z}$ takes zero values. The states of the interior of
permitted bands are organized in energy quadruplets, on certain
pairs of which $\mathcal{Z}$ takes nonzero values
$\pm\sqrt{P_{\mathcal{Z}}(E)}$. The diagonal components of
$\mathcal{Q}^2_{\pm}$ consist of differential operators of orders
$4r$ and $4(n-r)+2$. One of these two numbers is less than $2n+1$.
Suppose that it is the case of the operator $\mathcal{Q}^2_+$. Its
lower diagonal component $Q_+^{\dagger}Q_+$ satisfies a relation
$[H,Q_+^{\dagger}Q_+]=0$. According to the general theory of
finite-gap systems, the only operators commuting with $H$ of order
lower then $2n+1$ are polynomials in the Hamiltonian, and we
conclude that $Q_+^{\dagger}Q_+$ is such a polynomial, which has to
take zero values on $2r$ singlets belonging to $\mbox{Ker}\, Q_+$.
The same arguments hold for the upper component of the operator
$\mathcal{Q}^2_+$, and we find that
 \begin{equation}\label{Q+2}
     \mathcal{Q}^2_+=P_+(\mathcal{H})=\prod_{j=1}^{2r}(\mathcal{H}-E^{+}_j).
\end{equation}
With making use of (\ref{Z2}) and (\ref{Q+2}), we get also
\begin{equation}\label{Q-2}
    \mathcal{Q}^2_-=P_-(\mathcal{H})=
    \prod_{j=1}^{2(n-r)+1}(\mathcal{H}-E^{-}_j),
\end{equation}
where  $P_\pm(\mathcal{H})$ are positive-semidefinite  operators,
and $E^{\pm}$ are the energies of the corresponding band-edge states
annihilated by $\mathcal{Q}_{\pm}$. The eigenvalues of
$\mathcal{Q}_{\pm}$ are  $\pm \sqrt{P_{\pm}(E)}$, where the signs of
square roots are correlated with the  square root sign of
corresponding eigenvalue of $\mathcal{Z}$ in accordance with Eq.
(\ref{ZQQ}).

In addition to  the non-trivial integrals of motion $\mathcal{Z}$
and $\mathcal{Q}_{\pm}$,  the Hamiltonian $\mathcal{H}$ possesses
another triplet of trivial, mutually commuting, integrals
\begin{equation}\label{GGG}
    \Gamma_1=\sigma_3,\quad \Gamma_2=R,\quad \Gamma_3=R\sigma_3,
\end{equation}
which  satisfy the relations $\Gamma_i^2=1$, $i=1,2,3$,
$\Gamma_1\Gamma_2\Gamma_3=1$. Any of $\Gamma_i$ can be chosen as a
$\Z_2$-grading  operator $\Gamma_*$. Any of the non-trivial
integrals of motion either commutes or anti-commutes with any of the
trivial integrals. Fixing the grading operator, we classify any
nontrivial integral as a  bosonic or fermionic, while the
Hamiltonian and the trivial integrals (\ref{GGG}) are always
identified as bosonic operators. In correspondence with this
identification, a certain superalgebra is generated. For all the
three possible choices of the grading operator, one of the
nontrivial integrals plays the role of a bosonic, $\Z_2$-even
operator, while the other two integrals are classified as fermionic,
$\Z_2$-odd operators, see the Table below. We name this structure
the \textit{tri-supersymmetry}~\footnote{Such a structure  was
observed for the first time in the $N=2$ superextended Dirac delta
potential problem \cite{su22}, where the basic triplet of nontrivial
integrals has a completely different nature.  In particular, there
both supercharges of the hidden bosonized $N=2$ linear supersymmetry
of the form (\ref{linsusy}) are nonlocal operators, cf. (\ref{ZZH})
and (\ref{ZAR})}.

\begin{figure}[h!]
\begin{center}
\begin{tabular}{|l|l|l|l|}
\hline Grading operator & $\sigma _{3}$ & $R$ & $\sigma _{3}R$ \\
\hline
Bosonic integral & $\mathcal{Z}$ & $\mathcal{Q}_{+}$ & $\mathcal{Q}_{-}$ \\
\hline
Fermionic integrals & $\mathcal{Q}_{+}, \mathcal{Q}_{-}$ & $\mathcal{Z},~%
\mathcal{Q}_{-}$ & $\mathcal{Z},~\mathcal{Q}_{+}$ \\ \hline
\end{tabular}
\end{center}
 \label{tablina}
\end{figure}

The complete structure of the tri-supersymmetry will be described in
the next subsection. Here, let us choose $\Gamma_*=\sigma_3$ as the
grading operator, and discuss  the corresponding  nonlinear
supersymmetric subalgebra generated by the \emph{local} integrals of
motion, forgetting for the moment the nonlocal integral $R$.
Introduce the notation
\begin{equation}\label{N=4gen}
    \mathcal{Q}_{\pm}^{(1)}=\mathcal{Q}_{\pm},\quad
    \mathcal{Q}_{\pm}^{(2)}=i\sigma_3\mathcal{Q}_{\pm}.
\end{equation}
These fermionic supercharges together with bosonic operators
$\mathcal{Z}$ and $\mathcal{H}$ generate the  superalgebra
\begin{equation}\label{N=4Z++}
    \{\mathcal{Q}_{+}^{(a)},\mathcal{Q}_{+}^{(b)}\}=2\delta^{ab}P_+(\mathcal{H}),\quad
    \{\mathcal{Q}_{-}^{(a)},\mathcal{Q}_{-}^{(b)}\}=2\delta^{ab}P_-(\mathcal{H}),
\end{equation}
\begin{equation}\label{N=4Z+-}
     \{\mathcal{Q}_{+}^{(a)},\mathcal{Q}_{-}^{(b)}\}=2\delta^{ab}\mathcal{Z},
\end{equation}
\begin{equation}\label{ZHQ+-}
   [\mathcal{H},\mathcal{Q}_{\pm}^{(a)}]=[\mathcal{H},\mathcal{Z}]=
   [\mathcal{Z},\mathcal{Q}_{\pm}^{(a)}]=0.
\end{equation}
Superalgebra (\ref{N=4Z++}), (\ref{N=4Z+-}), (\ref{ZHQ+-}) is
identified as a centrally extended \emph{nonlinear} $N=4$
supersymmetry, in which $\mathcal{Z}$ plays a role of the bosonic
central charge \footnote{The basic structure of the algebra of
finite order differential operators of a form more general than
(\ref{N=4Z++})--(\ref{ZHQ+-}) was discussed by Andrianov and
Sokolov  \cite{AS2}, but outside the context of finite-gap
periodic systems and parity-even potentials. In comparison with
(\ref{N=4Z++})--(\ref{ZHQ+-}),  the superalgebraic structure of
Ref. \cite{AS2} includes some additional independent polynomial of
the Hamiltonian. As a consequence, instead of the relation
$\mathcal{Z}^2=P_+(\mathcal{H})P_-(\mathcal{H})$, which follows
from Eqs. (\ref{Z2})--(\ref{Q-2}) and reflects the nature and
peculiarities of the band structure \cite{Selfsusy}, its analog in
\cite{AS2} has a different form, see Eqs. (40), (41), (43) and
(46) there.}.

The supercharges $\mathcal{Q}_{+}^{(a)}$ annihilate a part of the
band-edge states organized in supersymmetric doublets, while
another part of supersymmetric doublets is annihilated by the
supercharges $\mathcal{Q}_{-}^{(a)}$. The band-edge states which
do not belong to the kernel of the supercharges
$\mathcal{Q}_{+}^{(a)}$ (or $\mathcal{Q}_{-}^{(a)}$) are
transformed (rotated) by these supercharges within the
corresponding supersymmetric doublet. The bosonic central charge
$\mathcal{Z}$ annihilates all the band-edge states.  So, we have
here the picture reminiscent somehow of the partial supersymmetry
breaking appearing in supersymmetric field theories with
BPS-monopoles~\cite{FauxSpec}.

\subsection{Tri-supersymmetry and $su(2|2)$}

Let us study the complete algebraic structure of the
tri-supersymmetry. To do this, consider the set of integrals of
motion created by  the multiplicative combinations of the trivial
integrals  with nontrivial ones,
 \begin{equation}\label{GGGG}
    \mathcal{H},\quad \Gamma_i, \quad
    \Gamma_\alpha\mathcal{Z},\quad \Gamma_\alpha\mathcal{Q}_+,\quad
    \Gamma_\alpha\mathcal{Q}_-,
\end{equation}
where $\alpha=0,1,2,3$, and by $\Gamma_0$ we denote a unit
two-dimensional matrix. Each of these integrals either commutes or
anticommutes with any of $\Gamma_i$ defined in (\ref{GGG}).
Identifying one of $\Gamma_i$ as the $\Z_2$-grading operator
$\Gamma_*$, we separate the set (\ref{GGGG}) into eight $\Z_2$-even
(bosonic) operators commuting with $\Gamma_*$, and eight $\Z_2$-odd
(fermionic) operators, which anticommute with $\Gamma_*$. Though
this  separation  depends on the choice of $\Gamma_*$, the
superalgebra in all three cases has, in fact, the same structure.

To reveal this common  superalgebraic structure for all the three
possible choices of the grading operator, we denote the fermionic
operators as ($F_1,\ldots,\, F_8 $). The set of bosonic operators we
write as ($\mathcal{H}$, $\Gamma_*$, $\Sigma_1$, $\Sigma_2$,
$B_1,\ldots, B_4$), where $\Sigma_{1,2}$ are two trivial integrals
from the set (\ref{GGG}) to be different from $\Gamma_*$. Denote by
$P_B(\mathcal{H})$ a universal  polynomial produced by the square of
any of the four integrals $B_a$, $B_a^2=P_B(\mathcal{H})$,
$a=1,\ldots,4$. Given $\Gamma_*$, we can separate the set of
fermionic operators in two subsets depending on the commutation
relations with the integrals $\Sigma_1$ and $\Sigma_2$. The subset
which commutes with $\Sigma _{2}$,  we label as $F_{\mu},
\mu=1,\ldots,4$, $[\Sigma _{2},F_{\mu}]=0$, and denote a universal
polynomial corresponding to a square of any of these fermionic
operators by $P_{22}$, $F_\mu^2=P_{22}$.  For the subset of
fermionic operators which commute with $\Sigma _{1}$, we put the
index $F_{\lambda},$ ${\lambda}=5,\ldots,8$, $[\Sigma
_{1},F_{\lambda}]=0$, and denote the analogous universal polynomial
by $P_{11}$, $F_\lambda^2=P_{11}$. Identifying  in the described way
the integrals $F_\mu$, $F_\lambda$, $B_a$, $\Sigma_1$ and
$\Sigma_2$, and computing directly all the (anti)commutators, we
find the superalgebra of tri-supersymmetry, which may be presented
in the same form modulo some special polynomials in dependence on
the choice of $\Gamma_*$. These additional polynomials we denote by
$P_{12}$, $P_{1B}$ and $P_{2B}$, where the subindex with two entries
indicates the origin of the (anti)commutation relation. Polynomial
$P_{12}$ comes from the anticommutators of $F_\lambda$ with $F_\mu$,
polynomial  $P_{2B}$  comes from the commutators between the
integrals $F_\mu$, ($[\Sigma _{2},F_{\mu}]=0$) and $B_a$, polynomial
$P_{1B}$ has analogous sense. Since the polynomials $P_{12}$,
$P_{1B}$ and $P_{2B}$ depend on the grading, their explicit form
together with explicit form of bosonic and fermionic operators for
all three choices of the grading operator are presented in Appendix
B, see Tables \ref{T1}, \ref{T2} and \ref{T3}. With the described
notations, the anti-commutation relations for fermionic operators,
and commutation relations between bosonic and fermionic operators
are presented in Tables \ref{T4} and \ref{T5}.

The identification of the complete superalgebra of the
tri-supersymmetry can be achieved now if we analyze the still
missing commutation relations between $\Z_2$-even  generators.
Introduce the following linear combinations of them,
\begin{equation}\label{K12}
     \mathcal{G}_{1}^{\left( \pm \right)
    }=\frac{1}{4} \left( B_{1}\pm B_{3}\right)=\frac{1}{2}B_1\Pi_\pm
    ,\quad
     \mathcal{G}_{2}^{\left( \pm \right) }=-\frac{1}{4}\left( B_{2}\pm
    B_{4}\right)=- \frac{1}{2}B_2\Pi_\pm ,
\end{equation}
\begin{equation}\label{K3}
    J_{3}^{\left( \pm \right) }=\frac{1}{4}\left( \Sigma _{1}\pm
    ~\Sigma _{2}\right) =\frac{1}{2}\Sigma _{1}\Pi_\pm ,
\end{equation}
where  $\Pi_\pm=\frac{1}{2}(1\pm \Gamma_*)$ are the projectors.
These operators satisfy the following algebra
\begin{equation}\label{K1K2}
    \left[ \mathcal{G}_{1}^{\left( \pm \right) },\mathcal{G}_{2}^{\left(
    \pm \right) }\right] =iJ_{3}^{\left( \pm \right)
    }P_{B}(\mathcal{H}),
\end{equation}
\begin{equation}\label{KaK3}
    \left[ J_{3}^{\left( \pm \right) },\mathcal{G}_{a}^{\left( \pm
    \right) }\right] =i\epsilon _{ab}\mathcal{G}_{b}^{\left( \pm \right)
    },\qquad a,b=1,2,
\end{equation}
\begin{equation}\label{KKi}
    [\mathcal{G}^{(+)}_a,\mathcal{G}^{(-)}_b]=
    [J_3^{(+)},\mathcal{G}^{(-)}_a]=[J_3^{(-)},\mathcal{G}^{(+)}_a]=
    [J_3^{(+)},J_3^{(-)}]=0,
\end{equation}
The commutation relations (\ref{K1K2})--(\ref{KKi}) correspond to
the direct sum of two deformed $su(2)$ algebras  in which
$\mathcal{H}$ plays a role of a multiplicative central charge. This
bosonic subalgebra is reminiscent of the nonlinear algebra satisfied
by the Laplace-Runge-Lenz and angular momentum vectors in the
quantum Kepler problem \cite{PasGal,Boer}.

It is known that in the case of the quantum Kepler problem, its
nonlinear symmetry algebra is reduced on the subspaces of fixed
energy $E<0$, $E=0$ and $E>0$ to the Lie algebras $so(4)$, $so(3,1)$
and $e(3)$, respectively, where $e(3)$ is the 3D Euclidean algebra.
Let us see what happens with our tri-supersymmetry under similar
reduction. First, consider any 4-fold degenerate energy level $E\neq
E_i$ corresponding to the interior part of any permitted band.
Rescaling the operators, $\mathcal{G}^\pm_a\rightarrow
J^{(\pm)}_a=\mathcal{G}^{(\pm)}_a/P_B(E)$, we find that together
with $J^{(\pm)}_3$ they generate the Lie algebra $su(2)\oplus
su(2)$. These operators satisfy the relations $J^{(+)}_i
J^{(+)}_i=\frac{3}{4}\Pi_+$, $J^{(-)}_i J^{(-)}_i=\frac{3}{4}\Pi_-$,
where the summation in $i=1,2,3$ is assumed. The two common
eigenstates of the Hamiltonian, with energy $E\ne E_i$, and of the
grading operator $\Gamma_*$, with eigenvalue $+1$  or $-1$, carry
the $1/2 \oplus 0$,  or $0\oplus 1/2$ representations of
$su(2)\oplus su(2)$, where the first (second) term corresponds to
the generators $J^{(+)}_i$ ($J^{(-)}_i$). The fermionic generators
mutually transform the states from the two eigenspaces of the
grading operator. In accordance with the total number of independent
fermionic generators, the energy subspace with $E\ne E_i$ carries an
irreducible representation of the $su(2|2)$ superunitary symmetry,
which is a supersymmetric extension of the bosonic symmetry
$u(1)\oplus su(2)\oplus su(2)$, where the $u(1)$ subalgebra is
generated by the grading operator, see Ref. \cite{Sen}. Having in
mind that the Hamiltonian appears in a generic form of the
superalgebra as a multiplicative central charge, we conclude that
the system possesses a nonlinear $su(2|2)$ superunitary symmetry in
the sense of Refs. \cite{nSUSY,MPhid,Boer}.

If we reduce our extended system to the subspace corresponding to
any doubly degenerate energy level $E_i$ corresponding to a doublet
of band-edge states, the bosonic part of the superalgebra is reduced
to the algebra $u(1)\oplus e(2)\oplus e(2)$, where the first term
corresponds to the integral $\Gamma_*$, while other two correspond
to the two copies of the 2D Euclidean algebras generated in the
eigensubspaces of $\Gamma_*$  by the rotation operators
$J^{(\pm)}_3$ and commuting translation generators
$\mathcal{G}^{(\pm)}_a$. Note that the supersymmetry of the form
similar to the present one reduced to a level $E_i$  was analyzed in
\cite{GT} in the context of spontaneous supersymmetry breaking in
3+1 dimensions.

\subsection{Self-isospectrality conjecture}

In the realm of supersymmetric quantum mechanics associated with
linear superalgebraic  structure, the complete isospectrality in the
non-periodic systems is related to the supersymmetry breaking, that
means that the doublet of the ground states is not annihilated by
supercharges. In \cite{dunfei}, Dunne and Feinberg considered
supersymmetric extensions of periodic potentials. They argued that
in contrary to the usual situation, the complete isospectrality of
super-partner Hamiltonians could appear without violation of the
supersymmetry. As one of the examples of the situation they
presented a one-gap Lam\'e Hamiltonian, where the super-symmetric
extension was provided by the first-order Darboux transformation
corresponding to the first order supercharges
$\mathcal{Q}_{-}^{(a)}$ defined by (\ref{N=4gen}). The super-partner
Hamiltonian showed to be the original one but displaced for a half
of the period. As we mentioned at the beginning, such a phenomenon
of a half-period displacement of super-partners was named in
\cite{dunfei} the self-isospectrality. We explained above how the
complete isospectrality emerges due to the tri-supersymmetry,
namely, its local part (\ref{N=4Z++})--(\ref{ZHQ+-}). In this
framework, the symmetries of the $N=2$ superextended one-gap Lam\'e
system have to be completed by adding two other supercharges
$\mathcal{Q}_{+}^{(a)}$ of order $2$ and a bosonic integral
$\mathcal{Z}$ of order $3$. The second order supercharges
$\mathcal{Q}_{-}^{(a)}$ do not annihilate the doublet of the ground
states, and the tri-supersymmetry is spontaneously partially broken.
As we showed, this turns out to be a general feature of the
tri-supersymmetric systems constucted by extension of a finite-gap
periodic system by means of a regular Crum-Darboux transformation.

We could ask, motivated by \cite{dunfei}, for the indications on the
self-isospectrality in the tri-supersymmetric extensions of the
finite-gap systems. In our current setting, the self-isospectrality
arises if the translation in the half-period $L$ provokes inversion
of the Wronskian,
\begin{equation}\label{isoWronsk}
    W^{\pm}(x+L)=\mathcal{C}^\pm\frac{1}{W^{\pm}(x)},
\end{equation}
where $\mathcal{C}^\pm$ are some nonzero constants. Indeed, such a
displacement produces changes in the sign for coefficient
functions $c_{1}^{\pm}(x)=-(\ln W^{\pm})'$ of the operators
$Q_{\pm}$, and therefore transforms the latters into their
conjugates,
\begin{equation}
    c_{1}^{\pm}(x+L)=-c_{1}^{\pm}(x), \qquad
    Q_{\pm}(x+L)=Q_{\pm}^{\dagger}(x).
\end{equation}
Making the translation for $L$ in the intertwining relations and
comparing the result with their conjugates, we reveal that
\begin{equation}
    \tilde{H}(x)=H(x+L),
\end{equation} and therefore the self-isospectrality does appear.

The construction considered in this section provides a receipt how
to get isospectral tri-super-symmetric partners for a given $n$-gap
Hamiltonian. We have seen how the partner Hamiltonian $\tilde{H}$ is
determined uniquely once we make a separation of the singlet states
into two disjoint families. There exist
$\sum_{k=0}^{n}\left(^n_k\right)=2^n$ distinct separations which
respect the rules explained in the subsection on the Darboux-Crum
transformations. Since one of them is trivial (includes all the
singlet states and corresponding integral $Z$ commutes with the
Hamiltonian $H$) we end up with $2^n-1$ tri-supersymmetric
isospectral extensions of the given $n$-gap system. All the
isospectral extensions can be obtained by successive first order
Darboux and second-order Crum transformations.

If antiperiodic singlet states are present in the spectrum, among
the possible separations of the singlet states there exists an
exceptional one, given by sorting out the singlets into mutually
orthogonal  families of periodic and antiperiodic states. Despite
the lack of the proof, we conjecture that this ``natural''
separation leads to the \textit{self-isospectral supersymmetry}
characterized by the partner Hamiltonian  $\tilde{H}$ to be the
original one but displaced in the half of the period.

Suppose an $n$-gap system $H$ with $n>1$ has antiperiodic singlet
states in its spectrum, and $\tilde{H}$ is a shifted for the
half-period Hamiltonian obtainable by the Darboux-Crum
transformation associated with the specified natural separation of
the singlets. Let $Q_{\pm}$ be the generators of the Crum-Darboux
transformation associated with a separation of singlets different
from the natural one. Then shifted for the half-period operators
$\tilde{Q}_{\pm}$ will generate a corresponding Crum-Darboux
transformation for the system $\tilde{H}$. In such a way we obtain a
new pair of self-isospectral systems $H_{Q}$ and $H_{\tilde{Q}}$:
\newsavebox{\iso}
    \savebox{\iso}{
    \rotatebox{0}{\scalebox{1}{
\begin{picture}(100,70)
\put(0,0){$\tilde{H}$} \put(0,60){$H$}
\put(90,0){$H_{\tilde{Q}}$} \put(90,60){$H_{Q}$}
\put(45,63){$^{Q_{\pm}}$} \put(45,3){$^{\tilde{Q}_{\pm}}$}
\put(7,30){$^{x\rightarrow x+L}$} \put(97,30){$^{x\rightarrow
x+L}$}
\put(5,58){\vector(0,-1){45}} \put(95,58){\vector(0,-1){45}}
\put(11,63){\vector(1,0){76}} \put(11,3){\vector(1,0){76}}
\end{picture}
   }}
    }
\begin{center}
\vspace{3mm}\usebox{\iso} \vspace{3mm}
\end{center}

\noindent Including $H$ and $\tilde{H}$, we can get $2^{n-1}$
distinct pairs of self-isospectral Hamiltonians. Starting with any
$n$-gap system system that has nonzero number of antiperiodic singlet
states, we would be able to construct $2^{n-1}$ extended
self-isospectral Hamitonians $\mathcal{H}$.

In the next section, our self-isospectrality conjecture will be
supported by the study of the tri-supersymmetry of the associated
Lam\'e equation.

\section{Associated Lam\'e equation and its isospectral extensiones}

We apply here a general theory developed in the previous sections
to a broad class of finite-gap systems described by the associated
Lam\'e equation. In particular we study the isospectral extension
based on the natural separation of the singlet states into
periodic and anti-periodic ones, and show that it leads to the
self-isospectral tri-supersymmetric systems. We provide an
explicit form of both the diagonal and non-diagonal integrals of
motion of the extended system. The examples of isospectral
extensions not possessing a property of self-isospectrality are
presented as well.

Associated Lam\'e equation is a two-parametric second order
differential equation of Fuchsian type with four singularities and
doubly-periodic coefficients,
\begin{equation}
     -\psi''-\left(C_m\mathrm{dn}^2x+C_l\frac{k'^2}{\mathrm{dn}^2x}+E\right)\psi=0,
     \label{lame}
\end{equation}
where $ C_m=m(m+1),\ \ C_l=l(l+1)$ are real numbers and
$\mathrm{dn}\, x\equiv \mathrm{dn}\,(x,k)$ is Jacobi elliptic
function with modular parameter $k\in (0,1)$;  $k'\in(0,1)$  is a
complementary modular parameter, $k'^2=1-k^2$. Lam\'e equation
($l=0$ case), obtained originally by separation of the Laplace
equation in elliptical coordinates, has been a subject of extensive
studies with use of both analytical
 \cite{Arscott,Erdelyi} and algebraical  \cite{AGI} methods.
Due to appealing properties of its solutions, the equation
(\ref{lame})  found the applications in diverse areas of physics.
In solid-state physics \cite{bel}, it represents a stationary
Schr\"odinger equation of a  model of one-dimensional crystal with
a more realistic potential than Kronig-Penney  or Scarf
potentials. This equation, especially its $l=0$ case, plays an
important role in many other fields of physics as well. For
instance, it appeared in some expansions of scattering amplitudes
\cite{MacWin}, in the study of bifurcations in chaotic hamiltonian
systems \cite{chaos}, it governs distance red-shift for partially
filled-beam optics in pressure-free FLRW cosmology
\cite{cosmology}, it was used in the study of static $SU(2)$ BPS
monopoles \cite{BPS} and kink solutions \cite{kink} in the field
theory.

\subsection{Construction of self-isospectral extension}

The spectrum of the one-dimensional periodic system governed by the
Hamiltonian operator corresponding to (\ref{lame})
\begin{equation}
    H^-_{m,l}=-D^2-C_m\mathrm{dn}^2x-C_l\frac{k'^2}{\mathrm{dn}^2x}
    \label{LAH}
\end{equation}
consists of the valence bands and the prohibited zones (gaps) which
alternate mutually until energy reaches a semi-infinite band of
conductance. Configuration of the spectral bands depends sensitively
on the constant parameters.  As long as $m$ and $l$ acquire integer
values, which we suppose to be the case from now on, the spectral
bands are arranged such that only finite number of gaps appear. The
period $2L$ of the potential with $C_m\neq C_l$ in (\ref{LAH}) is
equal to $2K$, where $K=\int_{0}^{\pi/2}(1-k^2\sin^2\phi
)^{-\frac{1}{2}}d\phi$ is the complete elliptic integral of the
first kind. The case $C_m=C_l$ corresponds to the Lam\'e system with
the same value of $C_m$ but $C_l=0$  and the period $2L=K$;  it is
discussed separately in the Appendix C. The independent change of
parameters $m\rightarrow -m-1$, $l\rightarrow -l-1$  leaves the
Hamiltonian (\ref{LAH}) invariant so that we can consider the case
$m>l\geq 0$ without the loss of generality. In this case the system
is $m$-gap.

To start on the construction of the tri-supersymmetric extension, we
focus to the separations of the band-edge states. As we announced,
the separation into the periodic and anti-periodic singlets will
be used here. Construction of the operators $Q_+$ and $Q_-$ associated
with any separation would require an explicit knowledge of the
band-edge states, i.e. an explicit solution of the stationary
Schr\"odinger equation. Fortunately, in the case of natural
separation this rather compelling work can be passed with the use of
peculiar properties of the model.

The present one-dimensional system is closely related to the
finite-dimensional representations of Lie algebra $sl(2,\R)$. The
Hamiltonian (\ref{LAH}) can be written as a second order polynomial
in generators of a finite-dimensional irreducible representation of
$sl(2,\R)$. This important feature underlies quasi-exact solvability
of the model, implying that a finite number of eigenstates
corresponding to band edges can be found by purely algebraic means
\cite{turbiner,ushveridze}. For integer values of $m$ and $l$,
$m>l\geq 0$, the space of $2m+1$ singlet states of the associated
Lam\'e system can be treated as a direct sum of two irreducible
non-unitary representations of $sl(2,\R)$ algebra of dimensions
$m-l$ (spin $j_{{}_-}=\frac{1}{2}(m-l-1)$) and $m+l+1$ (spin
$j_{{}_+}=\frac{1}{2}(m+l)$) \cite{qesfinkel, Ganguly}.

This fact is deeply related to the structure of the band-edge wave
functions of the system, $m+l+1$ of which  can be factorized
formally as
\begin{equation}
    \Psi_{\mu}=\mu \mathcal{F}_{\mu}(\xi),
    \label{psi1}
\end{equation}
whereas the remaining $m-l$ singlets acquire the form
\begin{equation}
    \Psi_{\nu}=\nu \mathcal{F}_{\nu}(\xi).
    \label{psi2}
\end{equation}
Here we introduced the functions
 $   \mu=\frac{\mathrm{cn}^{m+l}x}{\mathrm{dn}^lx},\quad
    \nu=\mathrm{cn}^{m-l-1}\,x \mathrm{dn} ^{l+1}\,x
 $    \label{gauge}
and a new variable
 $   \xi(x)=\frac{\mathrm{sn}\,x}{\mathrm{cn}\,x},
    \label{variable}$
that varies smoothly from $-\infty$ to $+\infty$ in the period
interval  $(-K,K)$. The functions $\mathcal{F}_{\mu}(\xi)$ and
$\mathcal{F}_{\nu}(\xi)$ are, in general, polynomials of order
$m+l$ and $m-l-1$ in $\xi$ and, as we will see, lie in vector
spaces of the irreducible representations of $sl(2,\mathbb{R})$ of
the dimensions $m+l+1$ and $m-l$.

As $\xi=\xi(x)$ is periodic, the factors  $\mu$ and $\nu$ dictate
periodicity or antiperiodicity of the eigenfunctions; for even
$m+l$ the wavefunctions (\ref{psi1})  are periodic while the
functions (\ref{psi2}) are anti-periodic.  For $m>l\geq 0$,
function $\mu$ has one node in the interval $(-K,K)$, while $\nu$
can have at most one node there. Being the polynomial in $\xi$,
the function $\mathcal{F}_{\mu}$ ($\mathcal{F}_{\nu}$) can acquire
at most $m+l$ ($m-l-1$) zeros in this interval. Combine these
facts with the general properties of periodic and anti-periodic
states of Hill's equation discussed in Section 2. The resulting
picture shows that starting with periodic ground state and
anti-periodic states at the edges of the first gap, the gaps with
periodic and anti-periodic states at their edges alter with energy
increasing until the $(m-l)$th gap is reached. For higher gaps,
all the remaining singlets at the edges are of the same nature as
edge states of the $(m-l)$th gap, periodic or anti-periodic with
even or odd number of nodes, see also ref. \cite{Selfsusy}.

To reveal the algebraic form of (\ref{LAH}), we shall recover how
the Hamiltonian acts on the ``dynamical'' part $\mathcal{F}_{\mu}$
($\mathcal{F}_{\nu}$) of the wavefunctions. Transforming out the
function $\mu$ and writing the result in the variable $\xi$ , we
obtain
\begin{eqnarray}
    h_{\mu}&=&(\mu)^{-1}H_{m,l}\mu=-\left(k'^2(T^+){}^2+(1+k'^2)
     (T^0){}^2+(T^-){}^2+k^2(l-m)T^0\right)\nonumber\\
      &=&-\left( 1+k^{\prime
    2}\xi ^{2}\right) \left( 1+\xi ^{2}\right) \frac{d^{2}}{d\xi
    ^{2}}+ \xi (2k^{\prime 2}(m+l-1)\xi ^{2}+2(m+lk^{\prime
    2}-1)+k^{2})\frac{d}{d\xi
    } \notag \\
    &&-k^{\prime 2}(m+l)(m+l-1)\xi ^{2}+const.
  \label{hT}
\end{eqnarray}
\noindent Here the operators
\begin{equation}
    T^{+}=\xi ^{2}\partial _{\xi }-(m+l)\xi ,\quad T^{0}=\xi \partial _{\xi }-\frac{m+l}{
    2},\quad T^{-}=\partial _{\xi } \, ,
    \label{sl2}
\end{equation}
\begin{equation}
    \left[ T^{+},T^{-}\right] =-2T^{0},~~~~\left[ T^{0},T^{\pm
    }\right] =\pm T^{\pm },
    \label{sl(2,R)}
\end{equation}
are the generators of irreducible representation of $sl(2,\R)$
acting on the vector space  spanned by monoms
$\{1,\xi,..,\xi^{m+l}\}$. Representation (\ref{sl2}) is specified
by the eigenvalue $j_{{}_+}(j_{{}_+}+1)$ of the Casimir
$C=-(T^0)^2+\frac{1}{2}(T^+T^-+T^-T^+)$ corresponding to
$sl(2,\R)$ spin $j_{{}_+}=\frac{1}{2}(m+l)$.

The other algebraic form of (\ref{LAH}), $h_{\nu}$, that acts on the
``dynamical" part  $\mathcal{F}_\nu$  of the wave functions
(\ref{psi2}), can be obtained by performing the gauge transformation
with the other common factor $\nu$. Alternatively, we can use the
apparent symmetry $\mu|_{l\rightarrow-l-1}=\nu$ and write down
$h_{\nu}$ immediately just by substituting $l\rightarrow-l-1$ in
(\ref{hT}) and (\ref{sl(2,R)}),
\begin{equation}
     h_{\nu}=h_{\mu}|_{l\rightarrow-l-1}=(\nu)^{-1}
     H_{m,l}^{-}\nu=-\left(k'^2(\tilde{T}^+)^2+(1+k'^2)
     (\tilde{T}^0)^2+(\tilde{T}^-)^2+k^2(-l-1-m)\tilde{T}^0\right).
\end{equation}
We denoted by $\tilde{T}^\rho=T^\rho|_{l\rightarrow-l-1}$,
$\rho=0,+,-$, the $sl(2,\R)$ generators of $(m-l)$-dimensional
representation, where $T^\rho$ are the generators  (\ref{sl2}) of
spin-$j_{{}_+}$ representation. Note that the ``effective"
Hamiltonian $h_{\mu}$ is Hermitian with respect to a scalar
product defined with a nontrivial weight,
$(f,g)=\int_{-\infty}^{\infty}f^*(\xi)g(\xi)
(1+k'^2\xi^2)^{-l+\frac{1}{2}}(1+\xi^2)^{-m+\frac{1}{2}}d\xi\,$;
the same is true for $h_{\nu}$ with the change $l\rightarrow
-l-1$.

Now the background of the natural separation of the singlet states
is clear.  The periodic and antiperiodic singlet states carry two
different irreducible representations of $sl(2,\R)$ of dimensions
$m+l+1$ and $m-l$. The number of the periodic and anti-periodic
singlet states depends on the values of $m$ and $l$, while their
total number $2m+1$ is fixed by the number of gaps $m$. For
instance, for $m=3,\ l=0$ there are $m-l=3$ periodic band-edge
states while for $m=3,\ l=1$ we have $m+l+1=5$ periodic singlet
states. To avoid possible confusions during the construction of the
supercharges, let us change the notation slightly. We will denote by
$X_{m,l}^{-}$ an operator which annihilates all the functions
(\ref{psi2}), and the operator annihilating all the states
(\ref{psi1}) will be $Y_{m,l}^{-}$.

First, let us consider eigenstates (\ref{psi1}) covered by the
$m+l+1$ dimensional representation of $sl(2,\mathbb{R})$. An
operator of the order $m+l+1$ which annihilates the representation
space spanned by the monoms $\{1,\xi,\ldots,\xi^{m+l}\}$ has the
following general form
\begin{equation}
     y_{m,l}^-=\alpha_{m,l}\partial_{\xi}^{m+l+1}.
     \label{yalpha}
\end{equation}
The function $\alpha_{m,l}$ is fixed uniquely as we require the
coefficient at $D^{m+l+1}$ of the operator
$
    Y_{m,l}^-=\mu\, y^-_{m,l}\frac{1}{\mu}\, \vert_{\xi=\xi(x)}
$ to be equal to one. It reads explicitly
$\alpha_{m,l}=\left(\frac{\mathrm{dn}\,x}{\mathrm{cn}^2x}\right)^{m+l+1}$.
We present below two equivalent forms of the operator $Y_{m,l}^-$.
The second, factorized expression, will be particularly helpful in
study of the limit case $k\rightarrow 1$. An explication how to get
it from (\ref{yalpha}) can be found in
\cite{tanakaorthogonalpolynomials},
\begin{eqnarray}
 Y_{m,l}^-&=&D^{m+l+1}+\sum_{j=1}^{m+l+1}c^Y_{j}D^{m+l+1-j}=
    \frac{\mathrm{dn}^{m+1}x}{\mathrm{cn}^{m+l+2}x}
    \left(\frac{\mathrm{cn}^2x}{\mathrm{dn}x}D\right)^{m+l+1}
    \frac{\mathrm{dn}^lx}{\mathrm{cn}^{m+l}x}\nonumber\\
    &=&\prod\limits_{j=-(m+l)/2}^{(m+l)/2}\left( D%
    +\left(\frac{k^2(m-l)\mathrm{cn}^2x}{2}-j(k'^2+
    \mathrm{dn}^2x)\right)\frac{\mathrm{sn}x}{\mathrm{cn}x\mathrm{dn}x}\right).
    \label{intY}
 \end{eqnarray}
The upper index of the ordered product corresponds to the first term
on the left side while the lower index denotes the last term on the
right side of the product. We can construct the operator $X_{m,l}^-$
in the same way or just by making the substitution $l\rightarrow
-l-1$ in (\ref{intY}) which interchanges considered algebraic
schemes. Explicitly, we get
\begin{eqnarray}
    X_{m,l}^-&=&D^{m-l}+\sum_{j=1}^{m-l}c^X_{j}D^{m-l-j}=
    \frac{\mathrm{dn}^{m+1}x}{\mathrm{cn}^{m-l+1}x}
    \left(\frac{\mathrm{cn}^2x}{\mathrm{dn}x}D\right)^{m-l}\frac{\mathrm{dn}^{-l-1}x}
    {\mathrm{cn}^{m-l-1}x}\nonumber\\&=&\prod\limits_{j=-(m-l-1)/2}^{(m-l-1)/2}\left(
    D
    +\left(\frac{k^2(m+l+1)\mathrm{cn}^2x}{2}-j(k'^2+\mathrm{dn}^2x)
    \right)\frac{\mathrm{sn}x}{\mathrm{cn}x\mathrm{dn}x}\right).
    \label{intX}
\end{eqnarray}
As we explained in the preceding section, the coefficients of the
second highest derivative of $X^-_{m,l}$ and $Y^-_{m,l}$ coincide
and enter the explicit construction of the superpartner Hamiltonian,
\begin{equation}
    c_1\equiv c_1^X=c_1^Y=-\frac{W_{m,l}^{\prime }}{W_{m,l}}=
    k^2\frac{(C_m-C_l)}{2}\frac{\mathrm{sn}x\mathrm{cn}x}{\mathrm{dn}x}=
    \frac{1}{2}{(m-l)(m+l+1)}k^{2}%
    \frac{\mathrm{cn}x\mathrm{sn}x}{\mathrm{dn}x}\, .\label{c1}
\end{equation}
The equality $c_1^X=c_1^Y$ reflects the coincidence of  the
Wronskians of the kernels of $X^-_{m,l}$ and $Y^-_{m,l}$ up to
inessential numerical factor related to the arbitrariness in
normalization of their zero modes. The essential part of these
Wronskians is given by the nodeless function
\begin{equation}\label{Wronkiy}
    W_{m,l}(x)=(\mathrm{dn}\, x)^{\frac{1}{2}(m-l)(m+l+1)},
\end{equation}
whose invariance with respect to the change $l\rightarrow-l-1$ just
reflects the indicated equality of the coefficients.

The Jacobi function property $\mathrm{dn}\, (x+K)=k'/\mathrm{dn}\,
x$ shows that the Wronskian $W_{m,l}(x)$ satisfies the relation
(\ref{isoWronsk}), and supports our conjecture that the natural
separation of the singlet states into periodic and anti-periodic
states results in the self-isospectral tri-supersymmetric system
with the partner Hamiltonian operator $\tilde{H}\equiv H^+_{m,l}$ to
be the original Hamiltonian translated for the half-period,
$
    H^+_{m,l}(x)=H^-_{m,l}(x+K).
$
Its explicit form is
\begin{equation}
    H^+_{m,l}=H_{m,l}^-+2c_1'=-
    D^2-C_l\mathrm{dn}^2x-C_m\frac{k'^2}{\mathrm{dn}^2x}.
\label{SUSYLAH}
\end{equation}

Recalling (\ref{ZZ}), the generator of the hidden bosonized
supersymmetry of the associated Lam\'e system (\ref{LAH}) acquires
the following factorized form

 \begin{eqnarray}
    Z_{m,l}^-&=&
    \frac{\mathrm{dn}^{-l}\,x}{\mathrm{cn}^{m-l+1}\,x}\left( \frac{
    \mathrm{cn}^{2}\,x}{\mathrm{dn}x}\frac{d}{dx}\right) ^{m-l}\left( \frac{
    \mathrm{dn}x}{\mathrm{cn}\,x}\right) ^{2m+1}\left( \frac{\mathrm{cn}^{2}\,x}{
    \mathrm{dn}x}\frac{d}{dx}\right) ^{m+l+1}\frac{\mathrm{dn}^{l}\,x}{\mathrm{cn
    }^{m+l}\,x}\\
    &=&\frac{\mathrm{dn}^{-l}x}{\,\mathrm{sn}^{m-l+1}\,x}\left( \frac{\mathrm{sn}
    ^{2}\,x}{\mathrm{dn}x}\frac{d}{dx}\right) ^{m-l}\left( \frac{\mathrm{dn}x}{
    \mathrm{sn}\,x}\right) ^{2m+1}\left( \frac{\mathrm{sn}^{2}\,x}{\mathrm{dn}x}
    \frac{d}{dx}\right)
    ^{m+l+1}\frac{\mathrm{dn}^{l}\,x}{\mathrm{sn}^{m+l}\,x}\, ,
\end{eqnarray}
where we used  alternative expressions
$X^-_{m,l}(x)=\frac{\mathrm{dn}^{m+1}x}
{\mathrm{sn}^{m-l+1}x}\left(\frac{\mathrm{sn}^2x}{\mathrm{dn}x}D\right)^{m-l}
\frac{\mathrm{dn}^{-l-1}x}{\mathrm{sn}^{m-l-1}x} $ and
$Y^-_{m,l}(x)=\frac{\mathrm{dn}^{m+1}x}
{\mathrm{sn}^{m+l+2}x}\left(\frac{\mathrm{sn}^2x}
{\mathrm{dn}x}D\right)^{m+l+1}\frac{\mathrm{dn}^lx}
{\mathrm{sn}^{m+l}x}$, obtained with the use of a specific identity
\begin{equation}\label{speciden}
    \frac{1}{\mathrm{sn}^{j+1}x}\left(\frac{\mathrm{sn}^2x}{\mathrm{dn}x}D\right)^{j}
    \frac{1}{\mathrm{sn}^{j-1}x}=\frac{1}{\mathrm{cn}^{j+1}x}
    \left(\frac{\mathrm{cn}^2x}{\mathrm{dn}x}D\right)^{j}\frac{1}{\mathrm{cn}^{j-1}x}.
\end{equation}
With making use of the same identity we can prove that
$Y^-_{m,l}(x+K)=(-1)^{m+l+1}(Y^-_{m,l}(x))^{\dagger}$  and
$X^-_{m,l}(x+K)=(-1)^{m-l}(X^-_{m,l}(x))^{\dagger}$. Finally, we
write down the obtained extended Hamiltonian $\mathcal{H}$ as well
as the Hermitian diagonal and antidiagonal integrals
 \begin{eqnarray}
    \mathcal{H}_{m,l}&=&\left(\begin{array}{cc} H^+_{m,l}&0\\0&H^-_{m,l}\end{array}\right)=
    \left(\begin{array}{cc} H^-_{m,l}(x+K)&0\\0&H^-_{m,l}(x)\end{array}\right),\label{HasL}\\
    \mathcal{Z}_{m,l}&=&i^{2m+1}\left(\begin{array}{cc} Z^+_{m,l}&0\\0&Z^-_{m,l}\end{array}\right)=i^{2m+1}
    \left(\begin{array}{cc} Z^-_{m,l}(x+K)&0\\0&Z^-_{m,l}(x)\end{array}\right),\label{ZasL}\\
    \mathcal{X}_{m,l}&=&i^{m-l}
    \left(\begin{array}{cc} 0&X^-_{m,l}\\X^+_{m,l}&0\end{array}\right)=i^{m-l}\left(\begin{array}{cc} 0&
    X^-_{m,l}(x)\\X^-_{m,l}(x+K)&0\end{array}\right),\label{XasL}\\
    \mathcal{Y}_{m,l}&=&i^{m+l+1}\left(\begin{array}{cc}
    0&Y^-_{m,l}\\Y^+_{m,l}&0\end{array}\right)=i^{m+l+1}
    \left(\begin{array}{cc}
    0&Y^-_{m,l}(x)\\Y^-_{m,l}(x+K)&0\end{array}\right),
    \label{matrixHXY}
\end{eqnarray}
which represent an explicit realization of (\ref{matHQZ}). In
accordance with the analysis of Section 3, the integrals
$\mathcal{Q}_{\pm}$ are identified with $\mathcal{X}_{m,l}$ and
$\mathcal{Y}_{m,l}$ in the following way:
$\mathcal{Q}_{-}=\mathcal{X}_{m,l}$ and
$\mathcal{Q}_{+}=\mathcal{Y}_{m,l}$ when $m-l$ is odd, and their
roles are interchanged when $m-l$ is even.

\subsection{Self-isospectral pairs and  ``superpotential"}

Let us indicate on an interesting representation of the
self-isospectral pairs of the Hamiltonians that generalizes a
representation $H_\pm=-D^2+W^2\pm W'$ of the super-partner
Hamiltonians in terms of the superpotential in the case of the usual
(linear) $N=2$ supersymmetry.

The self-isospectral pair (\ref{LAH}), (\ref{SUSYLAH}) can be
presented in the equivalent form
\begin{equation}
    H_{m,l}^{\pm}=-D^2+2\frac{C_m+C_l}{(C_m-C_l)^2}(\ln W_{m,l})'{}^2
    \pm (\ln W_{m,l})'' +(1+k'^{2})\frac{1}{2}(C_m+C_l),
    \label{hh}
\end{equation}
where $W_{m,l}$ is the Wronskian (\ref{Wronkiy}) corresponding to
the kernels of operators $X^-_{m.l}$ and $Y^-_{m,l}$. Let us denote
$C_+=\sqrt{\frac{1}{2}(C_m+C_l)}$, $C_-
=\frac{1}{2}(m-l)(m+l+1)=\frac{1}{2}(C_m-C_l)$, and define a
function
\begin{equation}\label{Wsusy}
    \mathcal{W}=-\left(\ln \mathrm{dn}^{C_+}x \right)^{\prime}.
\end{equation}
Then (\ref{hh}) can be rewritten equivalently as
\begin{equation}
     H^{\pm}_{m,l}-(1+k'^{2})C_+^2=-D^2+
     \mathcal{W}^{2}\pm \frac{C_-}{C_+}\mathcal{W}^{\prime }.
    \label{exact}
\end{equation}
Eq.  (\ref{exact}) is reminiscent of supersymmetric quantum
mechanics representation (\ref{HHEE}). This is not just a
coincidence. In the case $m-l=1$, the system $H^-_{m,m-1}$ is
characterized by the presence of only one periodic singlet
$\Psi_0=\mathrm{dn}^m\, x$, which is the ground state with energy
corresponding to a subtracted constant term on the left hand side of
Eq. (\ref{exact}). In this case the first order supercharge
$\mathcal{X}_{m,m-1}$ reduces to one of the first order supercharges
(\ref{Q12susy}) of $N=2$ supersymmetry, (\ref{Wsusy}) takes a form
of a usual representation of a superpotential in terms of the ground
state, and (\ref{exact}) reduces to (\ref{HHEE}).

There exists a simple generalization of a classical model for
supersymmetric quantum mechanics to the case of nonlinear
supersymmetry  of order $n>1$ \cite{MPhid}. It consists in the
change of the boson-fermion coupling term $\theta^+\theta^-W'$ in
classical Hamiltonian  for  $n\theta^+\theta^-W'$, where
$\theta^+\theta^-$ is a classical analog for $\sigma_3$ and
$\theta^\pm$ are Grassmann variables describing fermion degrees of
freedom. However, unlike the linear case $n=1$, for $n>1$ such a
generalization suffers a problem of the quantum anomaly, which can
be solved in a general form only for  $n=2$ \cite{KP}. One can show
that for $m-l=2$, when the supercharge $\mathcal{X}_{m,m-2}$ is the
differential operator of the second order, representation
(\ref{exact}) is in correspondence with  the solution of the quantum
anomaly problem for $n=2$.

Let us stress that for $m-l>1$ the argument of logarithm in the
definition of the superpotential-like function (\ref{Wsusy}) does
not correspond to a ground state of the system~\footnote{For the
explicit form of the ground states of the associated Lam\'e system
with some values of $m$ and $l$ see \cite{Ganguly}.}. It is just
the appropriately rescaled logarithmic derivative of the Wronskian
(\ref{Wronkiy}), $\mathcal{W}=-\frac{C_+}{C_-}(\ln W_{m,l})'$.

\subsection{Some examples}

Due to the general result presented in Section 3, there exist
$2^m-1$ isospectral extensions of an $m$-gap associated Lam\'e
Hamiltonian. The new systems can be obtained from the initial system
by sequent use of Darboux transformation and the transformation of
Crum of the second order. Moreover,  the complete set of $2^m$
isospectral systems can be sorted out into $2^{m-1}$
self-isospectral tri-supersymmetric pairs. We  present here an
example of Lam\'e system to illustrate this picture explicitly.

First, we explain briefly some subtleties related to the sequent use
of the transformation of Crum. Let us consider a two-gap associated
Lam\'e system represented by Hamiltonian $H$. There are three
admissible separations of the singlet states: sorting out the states
at the edges of the first gap, or of the second gap, or of both of
them.   Let us denote the supercharges associated with these
separations as $Q_{\pm,j},\ j=1,2,3$. Then we can construct three
isospectral super-partner Hamiltonians ${H}_{(j)}$ satisfying
 \begin{equation}
     H_{(j)}Q_{\pm,j}=Q_{\pm,j}H.
\end{equation}
We can repeat the procedure with anyone of the new systems,
obtaining another isospectral Hamiltonian $H_{(k,j)}$ ($j$ refers to
the system $H_{(j)}$, $k$ denotes the next choice of separation).
Although we can make this procedure repeatedly, it will generate
only limited number of new systems. This is due to simple rules
following the sequent use of transformations of Crum.

The first rule: when we choose sequently two identical separations,
we return to the initial system. In our two-gap setting, let us
start with the first separation and construct the new Hamiltonian
$H_1$ with the help of operator $Q_{+,1}$. The operator which
annihilates the states at the edges of the first gap of $H_1$
coincides with $Q_{+,1}^{\dagger}$. So, repeating the procedure with
$H_1$ with the same separation, we obtain a new Hamiltonian
$H_{(1,1)}$ which is related with initial $H$  by the intertwining
relation
$H_{(1,1)}Q_{\pm,1}^{\dagger}Q_{\pm,1}=Q_{\pm,1}^{\dagger}Q_{\pm,1}H$.
But the operator $Q_{\pm,1}^{\dagger}Q_{\pm,1}$ is a polynomial in
$H$, and we get $H_{(1,1)}=H$. In general, there holds
$H_{(j,j)}=H$.

The other rule tells that the choice of sequent separations is
``commutative''. Construct the Hamiltonian $H_1$ using the
supercharge $Q_{+,1}$. Then, choosing the second separation, we
construct $H_{(2,1)}$ with help of operator $\tilde{Q}_{+,2}$ which
annihilates the states at the edges of the second gap of the system
$H_1$. Hamiltonian $H_{(2,1)}$ satisfies a relation
$H_{(2,1)}\tilde{Q}_{\pm,2}Q_{\pm,1}=\tilde{Q}_{\pm,2}Q_{\pm,1}H$.
The operator $Q_{+,3}=\tilde{Q}_{\pm,2}Q_{\pm,1}$ is of the fourth
order and can be factorized in different ways.  For instance, its
alternative factorization is $Q_{+,3}=\tilde{Q}_{\pm,1}Q_{\pm,2}$,
that corresponds to the interchanged choices of the separations.
Speaking in general, there holds $H_{(k,j)}=H_{(j,k)}$.

Let us denote shortly $(j)$ (or $(j,k)$) a transformation of Crum of
the second (or of the fourth) order which annihilates singlet states
at the edges of the $j$-th gap (or of the both $j$-th and $k$-th
gaps). In this notation, Darboux transformation associated with the
ground state is represented by $(0)$. Coherently, we denote
$H_{(j)}$ or $H_{(j,k)}$ the Hamiltonians obtained by these
transformations. Then the rules for sequent use of Darboux-Crum
transformations can be depicted by the following schemes

\newsavebox{\comut}
    \savebox{\comut}{
    \rotatebox{0}{\scalebox{1}{
\begin{picture}(150,70)
\put(0,0){$H_{(j)}$} \put(0,60){$H$} \put(90,0){$H_{(j,k)}$}
\put(90,60){$H_{(k)}$} \put(48,61){$^{(k)}$} \put(48,3){$^{(k)}$}
\put(10,30){$^{(j)}$} \put(46,35){$^{(j,k)=(k,j)}$}
\put(96,30){$^{(j)}$} \put(5,58){\vector(0,-1){45}}
\put(93,58){\vector(0,-1){45}} \put(12,62){\vector(1,0){75}}
\put(16,3){\vector(1,0){72}} \put(12,58){\vector(3,-2){75}}
\end{picture}
   }}
    }

\newsavebox{\stejny}
    \savebox{\stejny}{
    \rotatebox{0}{\scalebox{1}{
\begin{picture}(160,70)
\put(0,0){$H_{}$} \put(90,0){${H}_{(j)}$}
\put(90,60){${H}_{(j,j)}=H$}
\put(45,3){$^{(j)}$} \put(97,30){$^{(j)}$} \put(30,30){$^{Id.}$}
\put(95,12){\vector(0,1){45}} \put(12,12){\vector(3,2){75}}
\put(11,3){\vector(1,0){76}}
\end{picture}
   }}
    }

\begin{center}
 \vspace{3mm}\usebox{\stejny} \usebox{\comut}

\vspace{3mm}
\end{center}
where $Id.$ represents an identity operator.

In the two-gap case, we can find three new isospectral
Hamiltonians in this way. Let us present isospectral
transformations of the $(m=3, l=0)$ Lam\'e Hamiltonian with their
relation to the original system. Seven new isospectral
Hamiltonains are found and four self-isospectral
 pairs can be formed. In the following scheme any of the new systems can
be reached by sequent application of Darboux  ($0$) or second-order
Crum's transformation $(k)$ on the initial Hamiltonian $H$. The
vertical lines correspond to transformation representing the natural
separation so that the Hamiltonians related by these lines form
self-isospectral pairs.

\newsavebox{\dvagap}
    \savebox{\dvagap}{
    \rotatebox{0}{\scalebox{1}{
\begin{picture}(190,70)
\put(30,30){$^{\mbox{two-gap}}$}
\put(0,0){$H_{(1)}$} \put(0,60){$H$} \put(90,0){$H_{(2)}$}
\put(90,60){$H_{(0)}$} \put(180,0){$H$} \put(180,60){$H_{(1)}$}
\put(5,58){\vector(0,-1){46}} \put(95,58){\vector(0,-1){46}}
\put(185,58){\vector(0,-1){46}}
\put(11,63){\vector(1,0){76}} \put(111,63){\vector(1,0){66}}
\put(21,3){\vector(1,0){66}} \put(111,3){\vector(1,0){66}}
\put(7,30){$^{(1)}$} \put(97,30){$^{(1)}$} \put(187,30){$^{(1)}$}
\put(45,63){$^{(0)}$} \put(135,63){$^{(2)}$}
\put(45,3){$^{(0)}$} \put(135,3){$^{(2)}$}
\end{picture}
   }}
    }

\newsavebox{\trigap}
    \savebox{\trigap}{
    \rotatebox{0}{\scalebox{1}{
\begin{picture}(370,90)
\put(33,30){$^{\mbox{three-gap}}$}
\put(0,0){$H_{(1,3)}$} \put(0,60){$H$} \put(90,0){$H_{(0)}$}
\put(90,60){$H_{(2)}$} \put(180,0){$H_{(2,3)}$}
\put(180,60){$H_{(1,2)}$} \put(270,0){$H_{(3)}$}
\put(270,60){$H_{(1)}$} \put(360,0){$H$} \put(360,60){$H_{(1,3)}$}
\put(45,63){$^{(2)}$} \put(135,63){$^{(1)}$}
\put(225,63){$^{(2)}$} \put(315,63){$^{(3)}$}
\put(45,3){$^{(2)}$} \put(135,3){$^{(1)}$} \put(225,3){$^{(2)}$}
\put(315,3){$^{(3)}$}
\put(7,30){$^{(1,3)}$} \put(97,30){$^{(1,3)}$}
\put(187,30){$^{(1,3)}$} \put(277,30){$^{(1,3)}$}
\put(367,30){$^{(1,3)}$}
\put(5,58){\vector(0,-1){46}} \put(95,58){\vector(0,-1){46}}
\put(185,58){\vector(0,-1){46}} \put(275,58){\vector(0,-1){46}}
\put(365,58){\vector(0,-1){46}}
\put(11,63){\vector(1,0){76}} \put(111,63){\vector(1,0){66}}
\put(208,63){\vector(1,0){60}} \put(291,63){\vector(1,0){66}}
\put(28,3){\vector(1,0){60}} \put(111,3){\vector(1,0){66}}
\put(208,3){\vector(1,0){60}} \put(291,3){\vector(1,0){66}}
\end{picture}
   }}
    }

\begin{center}
 \usebox{\trigap}
\end{center}
 \vspace{1mm}

\begin{figure}[h!]
\begin{center}
\includegraphics[scale=0.88]{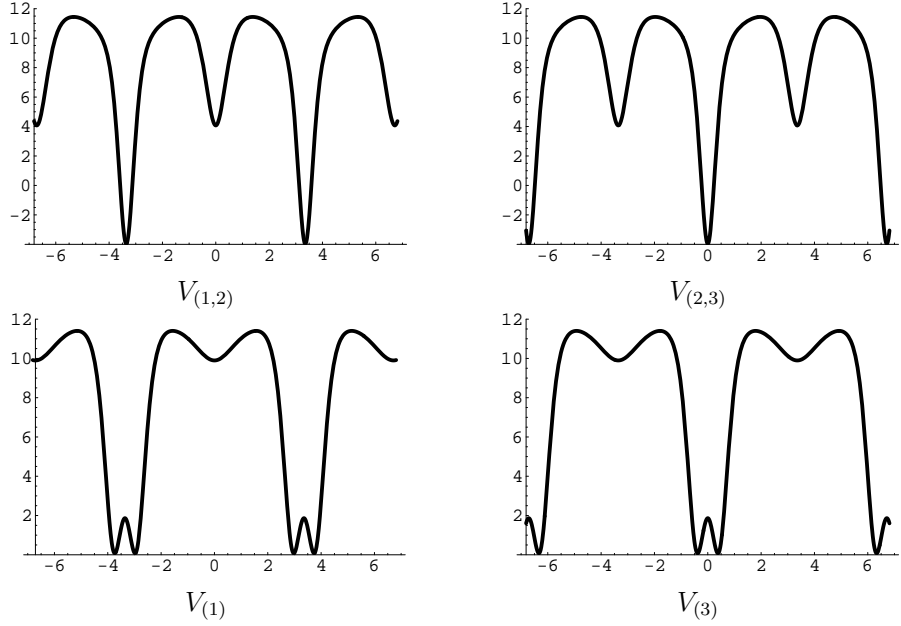}
\caption{The Hamiltonians $H_{(j,k)}=-D^2+V_{(j,k)}-12$ with the
plotted potentials are spectrally identical (see the last diagram in
sec. 4). The potentials on the left and right are mutually shifted
in half of the period. The modular parameter is set $k=0.99$.
\label{m=3iso}}
\end{center}
\end{figure}

As it can be observed in  Fig. \ref{m=3iso}, the nature of the
potentials of the obtained systems is distinct from the original
ones. However, the spectrum of the corresponding Hamiltonians is
completely identical. The potentials can be tuned with modular
parameter $k$ that broadens the applicability of these systems.
The infinite period limit of the self-isospectral extension of the
associated Lam\'e systems is discussed in detail in the
forthcoming section.

\section{Superextended P\"oschl-Teller system and
infinite period limit of tri-supersymmetry}

In this section we analyze in detail the infinite period limit of
the tri-supersymmetric self-isospectral extension
(\ref{HasL})--(\ref{matrixHXY}) of the associated Lam\'{e} system.
We explain how the structure of tri-supersymmetry is modified  and
study its implications, in particular the restoration of the
unbroken tri-supersymmetry.

Infinite period  limit corresponds to $k\rightarrow 1$
($k'\rightarrow 0$). In this limit the associated Lam\'e
Hamiltonian changes to the energy operator of the P\"oschl-Teller
system. The band structure transforms in the following way. The
states of the conduction band are transformed into the states of
the scattering sector of the spectrum, and the singlet edge-state
of the conduction band is transformed into the first (lowest)
singlet state of the continuous spectrum. The valence bands
shrink, two band-edge states corresponding to the same permitted
band converge smoothly in a unique wave function. This
wavefunction can be non-physical in one of the limit systems.

\subsection{Self-isospectral supersymmetry in the infinite-period limit}

When $k$ tends to one, the Jacobi elliptic functions cease to be
doubly periodic as their real period extends to infinity while the
complex period takes a finite value  $2iK^{\prime}=2i\pi$; they
transform into the hyperbolic functions,
   $  \mathrm{dn}\, x\rightarrow \mathrm{sech}\, x,~~\mathrm{cn}\, x
     \rightarrow \mathrm{sech}\, x,~~\mathrm{sn}\, x \rightarrow \mathrm{\tanh }\,x.$
In this limit, the superextended setting described by the two
mutually shifted periodic Hamiltonians (\ref{HasL}) acquires the
following form
\begin{equation}
    \mathcal{H}_{m,l}=\left(
    \begin{array}{cc}
    H_{m,l}^{+} & 0 \\
    0 & H_{m,l}^{-}
    \end{array}%
    \right)\underset{k=1}{\longrightarrow
    }\mathcal{H}_{m,l}^{PT}=\left(
    \begin{array}{cc}
    \hat{H}^+_{l} & 0 \\
    0 & \hat{H}^-_{m}
    \end{array}
    \right),
    \label{superPT}
\end{equation}
where the resulting  operators represent two systems with the
P\"oschl-Teller potential  of different interaction strengths
specified by the integers $m$ and $l$,
\begin{equation}
    \hat{H}^-_{m}=-\frac{d^{2}}{dx^{2}}-C_m\ \mathrm{sech}^{2}x,\quad
    \hat{H}^+_{l}=-\frac{d^{2}}{dx^{2}}-C_l\ \mathrm{sech}^{2}x.
    \label{PT12}
\end{equation}
The system keeps the parity-invariance,
$[R,\mathcal{H}_{m,l}^{PT}]=0$. As we deal with integer values of
 $m$ and $l$, Hamiltonians (\ref{PT12}) are reflectionless,
 i.e the transmission coefficients are equal to one.

\begin{figure}[h!]
\begin{center}
\includegraphics[scale=.88]{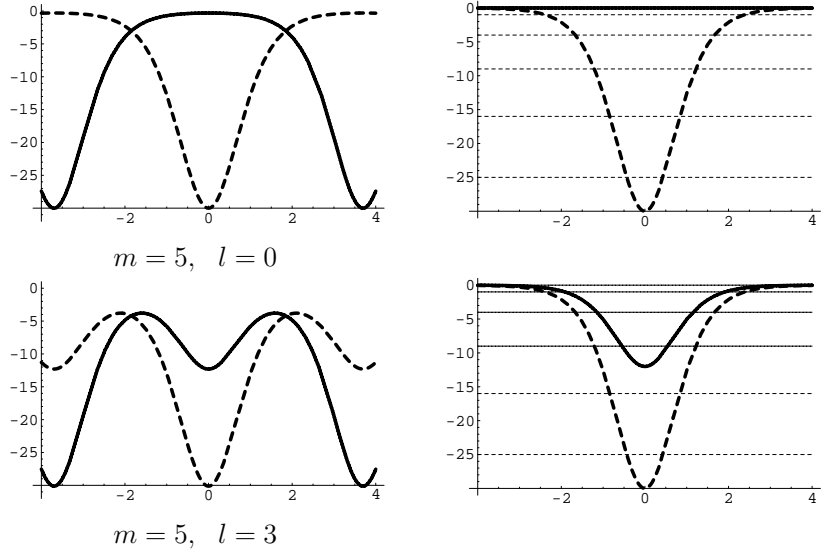}
\caption{Potentials on the left correspond to  $H^-_{m,l}$ (dashed
thick line) and $H^+_{m,l}$ (solid thick line); $k^2=0.99$. On the
right, the limit $k\rightarrow 1$ of these potential functions is
shown. The solid thin lines represent the shared bound-states and
the lowest scattering state. Dashed thin lines represent $m-l=2$
singlet states
\label{PTpictures} }
\end{center}
\end{figure}

The Hamiltonian $\hat{H}_{m}^-$ ($\hat{H}_{l}^+$) possesses $m+1$
($l+1$) singlet states, $m$ ($l$) of them are bound states, the
remaining one corresponds to the lowest state in the scattering
sector. Hamiltonians (\ref{PT12}) are almost-isospectral. Their
spectra coincide in the continuous part $E\in [0,\infty)$ and just
in $l+1$ singlet states. The Hamiltonian $\hat{H}_{m}^-$ has
additional $m-l$ lower energy levels. This indicates what happens
with the spectral structure of the extended Hamiltonian
$\mathcal{H}_{m,l}$ when we stretch the real period into the
infinity. The $2m+1$ band-edge states of $H_{m,l}^-$ transform
into $m+1$ physical states of $\hat{H}_{m}^-$, while in the case
of the super-partner system $H^{+}_{m,l}$ only $2l+1$ band-edge
states of highest excitations converge to the physical wave
functions, the rest is physically unacceptable. Thus the doublets
of band-edge states and quadruplets of the quasi-periodic states
change into $m-l$ singlets and $l+1$ doublet states,  and into the
quadruplets of the scattering states (see Fig.~\ref{PTpictures}).
The presence of the singlet states gives a taste of a different
nature of the tri-supersymmetry which we discuss in what follows.

The Hamiltonians (\ref{PT12}) admit the following representation

\begin{equation}
    \hat{H}^-_{m}=-\mathcal{D}_{-m}\mathcal{D}_m-m^2,\quad
     \hat{H}^+_{l}=-\mathcal{D}_{-l}\mathcal{D}_l-l^2,\label{HDD}
\end{equation}

\noindent where the definition and the basic properties of the
operator $\mathcal{D}_{n}$ are

\begin{equation} \mathcal{D}_{n}=D+n\tanh x, \quad
     \mathcal{D}^{\dagger}_{n}=-\mathcal{D}_{-n},
    \quad
    \mathcal{D}_{-n}\mathcal{D}_{n}=\mathcal{D}_{n+1}\mathcal{D}_{-n-1}+(2n+1).
    \label{D}
\end{equation}
We shall focus to the properties of tri-supersymmetry and of the
supercharges (\ref{ZasL})--(\ref{matrixHXY}) in particular. Taking
the limit $k=1$, the non-diagonal integrals (\ref{XasL}),
(\ref{matrixHXY}) transform as follows
\begin{equation}
 \mathcal{X}_{m,l}=i^{m-l}
    \left(\begin{array}{cc} 0&X^-_{m,l}\\X^+_{m,l}&0\end{array}\right)\underset{k=1}{\longrightarrow }
    \mathcal{X}_{m,l}^{PT}=i^{m-l}
    \left(\begin{array}{cc} 0&\hat{X}^-_{m,l}\\\hat{X}^+_{m,l}&0
    \end{array}\right),
    \label{maX}
\end{equation}
\begin{equation}
 \mathcal{Y}_{m,l}=i^{m+l+1}
    \left(\begin{array}{cc} 0&Y^-_{m,l}\\Y^+_{m,l}&0
    \end{array}\right)\underset{k=1}{\longrightarrow }
    \mathcal{Y}_{m,l}^{PT}=i^{m+l+1}
    \left(\begin{array}{cc} 0&\hat{Y}^-_{m,l}\\\hat{Y}^+_{m,l}&0\end{array}\right),
    \label{maY}
\end{equation}
where the non-diagonal components are
\begin{eqnarray}
     \hat{Y}^-_{m,l}&=&\mathcal{D}_{-l}\mathcal{D}_{-l+1}\ldots\mathcal{D}_{m-1}\mathcal{D}
        _{m},\qquad
     \hat{Y}^+_{m,l}=
     \mathcal{D}_{-m}\mathcal{D}_{-m+1}\ldots\mathcal{D}_{l-1}\mathcal{D}_{l},\\
    \hat{X}^-_{m,l}&=&\mathcal{D}_{l+1}\mathcal{D}_{l+2}\ldots\mathcal{D}_{m-1}\mathcal{D}
    _{m},\qquad
    \hat{X}^+_{m,l}=\mathcal{D}_{-m}
    \mathcal{D}_{-m+1}\ldots\mathcal{D}_{-l-2}\mathcal{D}_{-l-1},
\label{XYPT}
 \end{eqnarray}
and
 $   \hat{X}^+_{m,l}=(-1)^{m-l}(\hat{X}^-_{m,l})^{\dagger},$
$\hat{Y}^+_{m,l}=(-1)^{m+l+1} (\hat{Y}^-_{m,l})^{\dagger}.$ The
limit does not violate commutation relations (tri-supersymmetry is
maintained) and we have
\begin{equation}
    [\mathcal{H}_{m,l}^{PT},\mathcal{X}_{m,l}^{PT}]=[\mathcal{H}_{m,l}^{PT},
    \mathcal{Y}_{m,l}^{PT}]=[\mathcal{X}_{m,l}^{PT},\mathcal{Y}_{m,l}^{PT}]=0.\label{comutation}
\end{equation}

The components of the squares of the non-diagonal supercharges,
\begin{equation}
    (\mathcal{X}_{m,l}^{PT})^2=\left(
    \begin{array}{cc}
    \hat{X}^-_{m,l}\hat{X}^+_{m,l} &0 \\
    0 & \hat{X}^+_{m,l}\hat{X}^-_{m,l} \\
    \end{array}
    \right)
    ,\qquad
    (\mathcal{Y}_{m,l}^{PT})^2=\left(
    \begin{array}{cc}
    \hat{Y}^-_{m,l}\hat{Y}^+_{m,l} &0 \\
    0 & \hat{Y}^+_{m,l}\hat{Y}^-_{m,l} \\
    \end{array}%
    \right),
\label{polPT}
\end{equation}

\noindent correspond to the integrals of motion of individual
P\"oschl-Teller subsystems. The results of the previous section
suggest that these will be proportional to certain spectral type
polynomials in Hamiltonian. This is the case indeed. However, the
situation changes significantly comparing with the periodic system.
With the sequent use of the identity in (\ref{D}), we can derive

\begin{equation}
   (\mathcal{X}_{m,l}^{PT})^2=\prod_{j=0}^{m-l-1}(\mathcal{H}^{PT}_{m,l}-E_{m,j})=
   P^{PT}_X(\mathcal{H}^{PT}_{m,l}).
   \label{XXYY}
\end{equation}
Here $E_{m,j}=-(m-j)^2$, $j=0,\ldots,m-l-1,$ correspond to the
$m-l$ singlet states of the superextended system ($m-l$
bound-states correspond to the lowest energies of $\hat{H}^-_{m}$
or, equivalently to the $m-l$ nonphysical states of
$\hat{H}^+_{l}$). The square of the second non-diagonal
supercharge can be factorized with use of $P^{PT}_X$
\begin{equation}
    (\mathcal{Y}_{m,l}^{PT})^2=(\mathcal{H}^{PT}_{m,l}-E_{m,m})\prod_ {j=m-l}^{m-1}
    (\mathcal{H}^{PT}_{m,l}-E_{m,j})^2P^{PT}_X(\mathcal{H}^{PT}_{m,l}),
    \label{st}
\end{equation}

\noindent Apart from the roots shared with $P^{PT}_X$, there also
appear the $l+1$ doubly-degenerate energies of the superextended
system (\ref{superPT}). The $l$ bound-states energies
$E_{m,j}=-(m-j)^2$, $j=m-l,\ldots,m-1$ , are the double roots, and
the energy of the lowest state of the continuous spectrum,
$E_{m,m}=0$, is a simple root.

Considering the limit case of the diagonal supercharge
 (\ref{ZasL}), we encounter an interesting situation.
In \cite{FM} it was observed that for a \emph{reflectionless}
P\"oschl-Teller (PT) system, there exists a hidden bosonized
supersymmetry. If the system has $n$ bound states (and hence $n+1$
singlet states), there is a parity-odd integral of motion of order
$2n+1$,

\begin{equation}
    \mathcal{A}_{2n+1}=\mathcal{D}_{-n}\mathcal{D}_{-n+1}\ldots\mathcal{D}_{0}
    \ldots\mathcal{D}_{n-1}\mathcal{D}_{n},
    \label{bosoPT}
\end{equation}
which  annihilates all the singlet states and some
\emph{non-physical} states, whose origin was clarified in
\cite{Pecul} from the point of view of the Lam\'e equation and its
hidden bosonized supersymmetry~\footnote{Earlier, higher order
differential operators of this nature were discussed in the
context of supersymmetric quantum mechanics in  \cite{AS2,AS2*,
conformal}. However, their sense and the intimate relation with
the algebro-geometric potentials were not understood, see also
footnote 7.}. The subsystems $\hat{H}^-_{m}$ and $\hat{H}^+_{l}$
have odd-integrals of motion $\mathcal{A}_{2m+1}$ and
$\mathcal{A}_{2l+1}$ of orders $2m+1$ and $2l+1$, respectively. On
the other hand,  we know that the diagonal components of integral
$\mathcal{Z}^{PT}$ are the parity-odd integrals of order $2m+1$
for each subsystem. Particularly, $\hat{H}_l^+$ would have two
parity-odd symmetries of different orders. Let us explain the
picture and show how the integrals $\mathcal{A}_{2m+1}$ an
$\mathcal{A}_{2l+1}$ manifest their presence in the
tri-supersymmetric scheme.

In the limit
$\mathcal{Z}_{m,l}\underset{k=1}{\longrightarrow}\mathcal{Z}^{PT}_{m,l}$,
we can trace the presence of the integrals $\mathcal{A}_{2m+1}$ and
$\mathcal{A}_{2l+1}$ in the diagonal components

\begin{equation}
    Z^-_{m,l}\underset{k=1}{\longrightarrow}
    \hat{Z}^-_{m,l}=\hat{X}^+_{m,l}\hat{Y}^-_{m,l}=
    \mathcal{A}_{2m+1}=\mathcal{D}_{-m}
    \mathcal{D}_{-m+1}\ldots \mathcal{D}_{0}\ldots \mathcal{D}_{m-1}\mathcal{D}_{m},
    \label{Zm}
\end{equation}
\begin{equation}\label{Z+ml}
    Z^+_{m,l}\underset{k=1}{\longrightarrow}
    \hat{Z}^+_{m,l}=\hat{X}^-_{m,l}\hat{Y}^+_{m,l}=
    \hat{X}^-_{m,l}\hat{X}^{+}_{m,l}\mathcal{A}_{2l+1}.
\end{equation}
Each of these operators (it is worth to note again that they have
the same order) is the integral for the corresponding subsystem, and
together they annihilate the singlet and doublet states of the
super-extended system. $(\mathcal{Z}^{PT}_{m,l})^2$ produces a
polynomial of the form
\begin{equation}
    (\mathcal{Z}^{PT}_{m,l})^2=(\mathcal{H}^{PT}_{m,l}-
    E_{m,m})\prod_ {j=0}^{m-1}(\mathcal{H}^{PT}_{m,l}-E_{m,j})^2,
\label{spectralZPT}
\end{equation}
which can be related with a \emph{degenerate} spectral hyperelliptic
curve of genus $m$, in contrary
 to  the $m$-gap system (\ref{HasL}), whose  spectral polynomial
(\ref{Z2}) corresponds to a non-degenerate hyperelliptic curve of
the same genus. It reflects the fact that the band structure
disappeared; every two band-edge states of the same band transform
into a single bound state which ends up in the degeneracy in the
spectral polynomial. Accordingly, the degeneracy does not appear for
the lowest state of the continuous spectrum.

The components of the integral $\mathcal{Y}_{m,l}^{PT}$ can be
rewritten in the following way

\begin{eqnarray}
    \hat{Y}^-_{m,l}&=&\mathcal{A}_{2l+1}\mathcal{D}_{l+1}\ldots\mathcal{D}_{m-1}
    \mathcal{D}_{m}=\mathcal{A}_{2l+1}\hat{X}_{m}^{-}
    ,\qquad    \hat{Y}^+_{m,l}=\hat{X}_{l}^{+}\mathcal{A}_{2l+1}.
    \label{YcA}
\end{eqnarray}
The relation (\ref{st}) can be expressed also as
\begin{equation}
    (\mathcal{Y}_{m,l}^{PT})^2=\mathcal{A}^2_{2l+1}
    (\mathcal{H}^{PT}_{m,l})P^{PT}_X(\mathcal{H}^{PT}_{m,l}),
\end{equation}
where in correspondence with (\ref{bosoPT})
$\mathcal{A}_{2l+1}=\mathcal{D}_{-l}\mathcal{D}_{-l+1}\ldots
\mathcal{D}_{0}\ldots\mathcal{D}_{l-1}\mathcal{D}_{l}$. These
formulas provide an alternative insight into the kernel of the
supercharge $\mathcal{Y}_{m,l}^{PT} $ and its commutation relation
with $\mathcal{H}_{m,l}^{PT}$. The later one can be derived
independently just with use of the commutation relations
$[\mathcal{H}_{m,l}^{PT},\mathcal{X}_{m,l}^{PT}]=0$ and
$[\hat{H}_{l}^{+},\mathcal{A}_{2l+1}]=0$. In particular, we have
\begin{eqnarray}
  \hat{Y}_{m,l}^{-}\hat{H}_{m}^{-}&=&\mathcal{A}_{2l+1}
  \hat{X}_{m,l}^{-}\hat{H}_{m}^{-}=
    \mathcal{A}_{2l+1}\hat{H}_{l}^{+}\hat{X}_{m,l}^{-}=
    \hat{H}_{l}^{+}\mathcal{A}_{2l+1}\hat{X}_{m,l}^{-}=\hat{H}_{l}^{+}\hat{Y}_{m,l}^{-}
       \label{conju}
\end{eqnarray}
In the infinite-period limit, the tight relation of the non-diagonal
integrals (\ref{maX}) and (\ref{maY}) is manifested by means of the
parity-odd integral $\mathcal{A}_{2l+1}$ of $\hat{H}_{l}^{+}$, see
Eq. (\ref{YcA}),  which was not presented in the periodic case. As a
consequence, this integral appears  also in the structure of the
diagonal integral $\hat{Z}^+_{m,l}$, see (\ref{Z+ml}).

\subsection{Supercharges action and relation of nonphysical with physical solutions}

In the case of associated Lam\'e  system, the action of the
operators $\mathcal{X}^-_{m,l}$ and $\mathcal{Y}^-_{m,l}$ was
quite clear. Each of them annihilated two disjoint subsets of the
whole family of $2m+1$ singlet states of the Hamiltonian
$H_{m,l}^-$. In the limit case, the situation ceases to be so
transparent. The systems described by (\ref{PT12}) differ in the
number of the singlet states, the Hamiltonian $\hat{H}^{-}_{m}( \hat{H}^{+}_{l})$ has $m+1$ ($l+1$) bound-states. However, the
order of the supercharges is not affected by the limit. There
arises a natural question: what kind of functions is annihilated
additionally by the nontrivial integrals. We clarify here this
intricate situation.

We introduced the operator $\mathcal{D}_{n}$ (see (\ref{D})) which
proved to be useful in factorization of both the Hamiltonians
(\ref{HDD}) and the supercharges (\ref{maX}), (\ref{maY}). We can
interpret this operator as a Darboux tranformation which satisfies
\begin{equation}
     \mathcal{D}_{m}\hat{H}_{m}^-=\hat{H}_{m-1}^-\mathcal{D}_{m},
\end{equation}
where the new Hamiltonian $\hat{H}_{m-1}$ has $m-1$ bound states.
The operator $\mathcal{D}_{m}$ annihilates the ground state of
$\hat{H}^-_{m}$ so that the corresponding energy level is missing
in the spectrum of $\hat{H}_{m-1}$. The Hamiltonians $\hat{H}_{m}$
and $\hat{H}_{m-1}$ related by $\mathcal{D}_{m}$ are of the same
nature but with a shifted  parameter. This phenomenon, mediated by
Darboux transformation, is called a \emph{shape invariance}.

We can apply  the transformation of Darboux repeatedly, annihilating
the lowest bound state in each step. This procedure induces the
following sequence of Hamiltonians
\begin{equation}
\hat{H}_{m}^-\rightarrow
\hat{H}^{-}_{m-1}\rightarrow\hat{H}^{-}_{m-2}\rightarrow....
\rightarrow\hat{H}^-_{l+1}\rightarrow \hat{H}^-_{l}=\hat{H}^+_{l},
\label{sequence}
\end{equation}
The resulting operator which relates $\hat{H}_{m}^-$ and
$\hat{H}_{l}^+$ can be interpreted as a Crum-Darboux transformation
of order $m-l$. This transformation coincides with that produced by
the operator $\hat{X}_{m,l}^{-}$
\begin{equation}
     \hat{X}_{m,l}^{-}\hat{H}_{m}^-=\hat{H}^{+}_{l}\hat{X}_{m,l}^{-},\quad
     \hat{X}_{m,l}^{-}=\mathcal{D}_{l+1}\ldots\mathcal{D}_{m}.
    \label{PTDar}
\end{equation}

\noindent Therefore, $\hat{X}_{m,l}^{-}$ annihilates $m-l$ bound
states of $\hat{H}_{m}^{-}$. In the sense of the superextended
Hamiltonian $\mathcal{H}_{m,l}^{PT}$, these
bound-states are \emph{singlets}. The energy levels corresponding
 to the states annihilated by $\hat{X}_{m,l}^{-}$ are absent in the
spectrum of $\hat{H}_{l}^{+}$  (that makes the systems
almost-isospectral). Let us denote the corresponding $m-l$
bound-state energies of $\hat{H}^{-}_m$ as $E_{m}$. We denote
$l+1$ remaining energies of singlet states of $\hat{H}_{m}^{-}$ as
$E_{l}$. These $l+1$ levels are shared by the singlet states of
$\hat{H}_{l}^{+}$. So, the energies of singlet states of
$\hat{H}_{m}^{-}$ are formed by $E_m$ and $E_{l}$.

The operator
$\hat{X}^{+}_{m,l}$ annihilates a non-physical eigenstate
 of $\hat{H}_{l}^{+}$ corresponding to the energy $E_m$
$
    \hat{X}^{+}_{m,l}\tilde{\psi}_m=0,
$
$
    \hat{H}_{l}^{+}\tilde{\psi}_m=E_{m}\tilde{\psi}_m
$
Acting with the operator $\hat{X}^{+}_{m,l}$ on the $l+1$
physical states ($l$ bound states and the lowest state of the
continuous spectrum) $\tilde{\psi}_{l}$ of $\hat{H}_{l}^{+}$, $
\hat{H}_{l}^{+}\tilde{\psi}_{l}=E_{l}\tilde{\psi}_{l}$, we get the
bound states of $\hat{H}_{m}^{-}$ corresponding to the energy
$E_l$
$
    \hat{X}^{+}_{m,l}\tilde{\psi}_{l}=\psi_{l},
$
$
    \hat{H}_{m}^{-}\psi_{l}=E_l\psi_{l}.
$ The operator $\hat{Y}^{+}_{m,l}$  annihilates the physical
eigenstates $\tilde{\psi}_{l}$,
$
    \hat{Y}^{+}_{m,l}\tilde{\psi}_{l}=0,
$ and also annihilates the second, non-physical, solution
$\tilde{\eta}_m$ of  the Hamiltonian $\hat{H}_{l}^{+}$
corresponding to the eigenvalue $E_{m}$,
$
    \hat{Y}^{+}_{m,l}\tilde{\eta}_m=0,
$
$
    \hat{H}_{l}^{+}\tilde{\eta}_m=E_{m}\tilde{\eta}_m,
$
$
    \tilde{\eta}_m=\tilde{\psi}_m\int^x\tilde{\psi}_m^{-2}dx.
$
The function $\tilde{\eta}_m$ is mapped by $\hat{X}^{+}_{m,l}$
to a physical state $\psi_{m}$ of $\hat{H}_{m}^{-}$ with energy
$E_m$. The same result is obtained when we act by
$\hat{Y}^{+}_{m,l}$ on the state $\tilde{\psi}_m\in \mathrm{Ker}\,
\hat{X}^{+}_{m,l}$,
$
    \hat{Y}^{+}_{m,l}\tilde{\psi}_m=\psi_{m},
$
$
    \hat{X}^{+}_{m,l}\tilde{\eta}_m\propto\psi_{m},
$
$
    \hat{H}_{m}^{-}\psi_{m}=E_{m}\psi_{m}.
$

To get an insight into the kernel of the operator $\hat{Y}_{m,l}^-$,
it is convenient to rewrite this operator in the factorized form
\begin{equation}
     \hat{Y}_{m,l}^{-} =\mathcal{D}_{-l}\mathcal{D}_{-l+1}\ldots\mathcal{D}_{-1}
    \underbrace{\mathcal{D}_{0}\mathcal{D}_{1}\ldots
    \mathcal{D}_{m-1}\mathcal{D}_{m}}_{\text{annihilates
singlets}}.
\end{equation}
The indicated right part of the operator annihilates all the $m+1$
singlet states of $\hat{H}_{m}^{-}$. Due to the remaining part,
there are additional annihilated  $l$ functions $\phi_j$, $\
j=m-l,\ldots,m-1$. These additional functions need not to be
solutions of the Schr\"odinger equation corresponding to
$\hat{H}^{-}_{m}$. Indeed, due to representation
\begin{equation}
\hat{Y}_{m,l}^{+}\hat{Y}_{m,l}^{-}=(\hat{H}^{-}_{m}-E_{m,m})\prod_
{j=m-l}^{m-1}
    (\hat{H}^{-}_{m}-E_{m,j})^2P^{PT}_X(\hat{H}^{-}_{m}),
\label{spectralYPT}
\end{equation}
 the functions $\phi_j$ satisfy
$(\hat{H}^{-}_{m}-E_{m,j})\phi_j=\psi_j,$
$(\hat{H}^{-}_{m}-E_{m,j})\psi_j=0.$ Hence, the Hamiltonian
$\hat{H}_{m}^-$ restricted on the kernel of $\hat{Y}_{m,l}^{-}$
contains Jordan blocks associated with the energies $E_l$. Due to
the similar structure of the spectral polynomial of
$\hat{Z}_{m,l}^{-}$ (\ref{spectralZPT}), the non-physical states
annihilated by the supercharge of the hidden bosonized
supersymmetry are of the same nature.

It is instructive to discuss the case of $l=0$ in detail. In this
case $\hat{H}_{0}^{+}$ corresponds to the free particle. The
commutation relations (\ref{comutation}) tells that the free
particle energy operator is related with $\hat{H}_m^{-}$ via the
following intertwining relations
\begin{equation}
\hat{X}^{+}_{m,0}\hat{H}_0^{+}=\hat{H}_m^{-}\hat{X}^{+}_{m,0},\ \
\hat{Y}^{+}_{m,0}\hat{H}_0^{+}=\hat{H}_m^{-}\hat{Y}^{+}_{m,0},
\end{equation}
\begin{equation}
\hat{X}^{-}_{m,0}\hat{H}_m^{-}=\hat{H}_0^{+}\hat{X}^{-}_{m,0},\ \
\hat{Y}^{-}_{m,0}\hat{H}_m^{-}=\hat{H}_0^{+}\hat{Y}^{-}_{m,0}.
\label{infree}
\end{equation}

\noindent
The first relation of (\ref{infree}) is mediated by $
\hat{X}^-_{m,0}=\mathcal{D}_{1}...\mathcal{D}_{m}$. Keeping in mind
(\ref{PTDar}), the supercharge $\hat{X}^{-}_{m,0}$ annihilates all
the bound-states of $\hat{H}_{m}^{-}$. In the second relation of
(\ref{infree}), the Hamiltonians are intertwined by
$\hat{Y}^{-}_{m,0}=\mathcal{D}_0\mathcal{D}_1...\mathcal{D}_m$.
Apparently, this operator makes the same job as $\hat{X}^{-}_{m,0}$
but $\hat{Y}^{-}_{m,0}$  annihilates additionally the first
scattering state of $\hat{H}_{m}^{-}$.

The operator $\hat{X}^{-}_{m,0}$ transforms the lowest scattering
state of $\hat{H}_{m}^{-}$ into a constant function, the scattering
state of a free particle corresponding to the lowest energy. This
function is annihilated by $\hat{Y}^{+}_{m,0}$ while applying
$\hat{X}^{+}_{m,0}$ we get the initial first scattering state of
$\hat{H}_{m}^{-}$. In general, the operators $\hat{X}^{+}_{m,0}$ and
$\hat{Y}^{+}_{m,0}$ transform solutions of Schr\"odinger equation
corresponding to the free particle $H_{0}^{+}$ into the (formal)
eigenstates of $\hat{H}_{m}^{-}$. They can be employed in
reconstruction of the scattering states of the Hamiltonian
$\hat{H}_m^{-}$ from the plane wave states of a free particle,

\begin{eqnarray}
    \psi_{\kappa}^{\pm}&=&\hat{X}_{m,0}^{+}e^{\pm i\kappa
    x}=\mathcal{D}_{-m}\mathcal{D}_{-m+1}\ldots
    \mathcal{D}_{-2}\mathcal{D}_{-1}e^{\pm i\kappa x} \nonumber\\
    &\propto&\hat{Y}_{m,0}^{+}e^{\pm i\kappa
    x}=\mathcal{D}_{-m}\mathcal{D}_{-m+1}\ldots
    \mathcal{D}_{-1}\mathcal{D}_{0}e^{\pm i\kappa x},
\end{eqnarray}

\noindent where $\tilde{\psi}_{\kappa}^{\pm}=e^{\pm i\kappa x}$
satisfies

\begin{eqnarray}
    \hat{H}_{0}^{+}\tilde{\psi}_{\kappa}^{\pm}&=&E_\kappa
    \tilde{\psi}_{\kappa}^{\pm},\quad \hat{H}_{m}^{-}\psi_{\kappa}^{\pm}=E_\kappa
    \psi_{\kappa}^{\pm},\quad E_k=\kappa^2.
\end{eqnarray}

Let us summarize the obtained results. Extending the real period of
the self-isospectral extension of associated Lam\'e Hamiltonian to
infinity, the associated superalgebraic structure was modified. The
squares of the supercharges $\mathcal{Z}_{m,l}^{PT}$ and
$\mathcal{Y}^{PT}_{m,l}$ turned out to be degenerated polynomials in
$\mathcal{H}_{m,l}^{PT}$. This  modifies the structure of the
underlying superalgebra of the bosonic operators
$\mathcal{G}_a^{(\pm)}$ in the dependence on a chosen grading
operator $\Gamma_*$. Comparing with the periodic case, the bosonic
operators form the algebra  $u(1)\oplus e(2) \oplus e(2)$ for the
singlet energy levels. Recall that the periodic \emph{extended}
tri-supersymmetric system  does not have singlet states in its
spectrum.

The operators $\hat{X}^{-}_{m,l}$ and $\hat{Y}^{-}_{m,l}$
annihilate the $m-l$ lowest bound states of the Hamiltonian
$\hat{H}_{m}^-$. The eigenvectors of $\hat{H}_{l}^+$ corresponding
to these energies cease to be physically acceptable. Consequently,
isospectrality of the initial system is broken. Speaking in terms
of the extended system, the supercharge $\mathcal{X}^{PT}_{m,l}$
annihilates the singlet states. The operator $\mathcal{Y}^{PT}_m$
annihilates both doublets and singlets which are annihilated by
the diagonal operator $\mathcal{Z}^{PT}_m$ as well. From this
point of view, the spontaneously (or, dynamically) partially
broken tri-supersymmetry of the periodic system is recovered in
the infinite period limit.

\section{Concluding remarks and outlook}

In the particular case of associated Lam\'e systems, the results
of the present paper should be understood in a broader context of
the existing literature. Dunne and Feinberg considered the class
$l=m-1$ of associated Lam\'e Hamiltonians (\ref{LAH}) as an
example of the self-isospectral extension provided by Darboux
transformation \cite{dunfei}. Khare and Shukhatme \cite{noiso}
found that this transformation provides a self-isospectral
extension of pure Lam\'e systems just in the one-gap case while
for the other setting the extension proved to be of a completely
different nature. On the other hand, Fern\'andez \textit{et al}
revealed self-isospectrality of two-gap Lam\'e Hamiltonian when
the second-order transformation was applied \cite{FerNegNie}.

In the light of the presented results, we can understand those
findings just as pieces of the mosaic, which was fully unfolded by
the structure of the tri-supersymmetry and especially by the
self-isospectral supersymmetry of the associated Lam\'e system. In
particular, the system considered by Dunne and Feinberg is  the
self-isospectral extension $\mathcal{H}_{m,m-1}$ of the associated
Lam\'e Hamiltonian, see (\ref{matrixHXY}). Besides the first-order
supercharge $\mathcal{X}_{m,m-1}$, the list of its local integrals
of motions should be completed by the other non-diagonal
supercharge $\mathcal{Y}_{m,m-1}$ and diagonal integral
$\mathcal{Z}_{m,m-1}$ which  plays the role of the central charge of
the resulting  extended $N=4$ nonlinear supersymmetry. Although both
$\mathcal{X}_{m,m-1}$ and $\mathcal{Z}_{m,m-1}$ annihilate the
doublet of ground states, the tri-supersymmetry is spontaneously
partially broken since the doublet of ground states does not vanish
under the action of the supercharge $\mathcal{Y}_{m,m-1}$. This
suggests that the  supersymmetry breaking should be analyzed having
in mind the complete set of nontrivial local integrals, which are
$\mathcal{Z}$, $\mathcal{Q}_{+}^{(a)}$ and $\mathcal{ Q}_{-}^{(a)}$
in the case of the studied general class of finite-gap systems.

On our way to the presented results we left untouched various
appealing questions and problems. For instance, the
self-isospectrality conjecture could be tested on the finite-gap
systems with missing anti-periodic states. Since these should be
prevented from the self-isospectral extensions, the structure of the
tri-supersymmetry could exhibit peculiarities   in this case.
Besides, the exact proof of the conjecture should be provided.

The infinite-period limit could be an effective technique in
production of the tri-supersymmetric systems with non-periodic
potentials. In the limit case of the self-isospectral extension of
the associated Lam\'e Hamiltonian, the isospectrality was broken
followed by the recovery of the exact tri-supersymmetry. There
appears a natural question whether this is the common feature or
there exist isospectral extensions of non-periodic systems with
broken tri-supersymmetry. The limit of other isospectral
extensions of associated Lam\'e system could provide an insight
into the general situation. The relation of the tri-supersymmetry
and the representations of Lie algebras might give an interesting
insight into involved physical models as well.

Our construction of the tri-supersymmetric extensions was based on
the specific factorization of the odd-order integral of motion.
Relaxing the smoothness of the potential, the formal construction
should be applicable on the broad family of algebro-geometric
potentials where the presence of the parity-odd diagonal integral
$\mathcal{Z}$ is guaranteed. It is worth to mention the
Treibich-Verdier family of potentials \cite{TV} in this context.
Besides the associated Lam\'e systems, this family contains
singular potentials, which could be convenient examples to study
the tri-supersymmetry in singular systems.

Regular Crum-Darboux transformations with zero modes in the
prohibited  bands can produce self-isospectral potentials with a
generic shift of the  coordinate,  or superpartners with
periodicity defects. The particular results of this type were
obtained in \cite{FerMiletal,sgnn,FMRS,FerGan1,FerGan2}
 with making use of the first- and the second-order
 transformations applied to one- and two-gap Lam\'e equation.
 It would be interesting to analyse such a class of systems
 on the presence of the tri-supersymmetric structure.

The revealed supersymmetric structure was based on the internal
properties of the integral of motion $Z$, related with the KdV
hierarchy. This indication of the tri-supersymmetry and
self-isospectrality in the context of nonlinear integrable systems
should be followed and analyzed. A special attention should be
paid to possible manifestations of the tri-supersymmetry in
physical systems \cite{Selfsusy}.

\bigskip

The work has been partially supported  by CONICYT, DICYT (USACH) and
by FONDECYT under grants 1050001 and 3085013. We are grateful to V.
Enolskii, V. Spiridonov, A. Treibich, R. Weikard and A. Zabrodin for
valuable communications. Our special thanks are to B. Dubrovin for
many detailed explanations on the theory of finite-gap systems.

\setcounter{equation}{0}
\renewcommand{\theequation}{A.\arabic{equation}}

\setcounter{equation}{0}
\renewcommand{\theequation}{A.\arabic{equation}}

\section*{Appendix A}

Higher-order differential operators play the key role in the
construction of the tri-supersymmetry since they mediate
intertwining of the super-partner Hamiltonians. As we explained in
the section on the Crum-Darboux transformations, properties of these
operators are determining for the physical characteristics of the
superpartner systems. We present here a short resume of the relevant
facts referring for the details to \cite{Ince}, \cite{DarSol}.

Consider a differential operator of order $n$ which annihilates $n$
functions $\psi_i, \; i=1,\ldots,n$,

\begin{equation}
    A_n=D^{n}+\sum_{j=1}^{n}c_{j}^{A}(x)D^{n-j}, \quad A_n\psi_i=0,
    \quad
     i=1,...,n.
     \label{Q}
\end{equation}

\noindent Its coefficients are determined by the functions
$\psi_i$. For instance, the coefficient $c_{1}^{A}(x)$ can be
given in terms of the Wronskian of the $n$ functions $\psi_i$,
$c_{1}^{A}(x)=-\frac{d}{dx}\ln W(\psi _{1},\ldots,\psi_{n}),$
where
 $   W(\psi _{1},\ldots,\psi _{n})=W=\det B, \quad
    B_{i,j}=\frac{d^{j-1}\psi _{i}}{dx^{j-1}},
    \quad i,j,=1,\ldots,n.$
This is in accordance with the general form for the coefficients
 $   c_{j}^{A}(x)=-\frac{W_{j}}{W}
     , \quad j=1,\ldots,n,$
where $W_{j}$ is the determinant of the matrix  $B$ modified by
replacing the line $\psi _{1}^{(n-j)},\ldots,\psi _{n}^{(n-j)}$ by
$\psi _{1}^{(n)},\ldots,\psi _{n}^{(n)}$. In this notation,
$W_0\equiv W$.

The operator $A_n$ can be factorized in terms of the first order
differential operators. There follow equivalent representations of
$A_n$ which provide a better insight into the properties of the
operator, see \cite{Ince},

\begin{equation}
    A_n=(-1)^n\frac{W_{n}}{W_{n-1}}D\frac{W_{n-1}^{2}}{W_{n}W_{n-2}}
    D\ldots D\frac{W_{1}^{2}}{W_{2}W_{0}}D\frac{ W_{0}}{W_{1}},
\end{equation}
We can write equivalently
\begin{equation}
    A_n=L_{n}L_{n-1}\ldots L_{2}L_{1}, \quad
    L_{j}=D-\alpha_j, \quad
    \alpha_j=\frac{d}{dx}\ln \frac{W_j}{W_{j-1}},\quad j=1,\ldots,n.
\end{equation}
The operator can also be expressed as a determinant
\begin{equation}
A_n=W^{-1}(\psi_1,\ldots, \psi_n)\left\vert
\begin{array}{ccccc}
    \psi _{1} & \psi _{2} & \cdots  & \psi _{n} & 1 \\
    \psi _{1}^{\prime } & \psi _{2}^{\prime } & \cdots  & \psi
    _{n}^{\prime }
    & D \\
    \vdots  & \vdots & \ddots  & \vdots  & \vdots  \\
    \psi _{1}^{(n-1)} & \psi _{2}^{(n-1)} & \cdots  & \psi _{n}^{(n-1)} & D^{n-1} \\
    \psi _{1}^{(n)} & \psi _{2}^{(n)} & \cdots  & \psi _{n}^{(n)} & D^n
\end{array}
\right\vert ,
\end{equation}
where the multiplicative factor fixes the coefficient of $D^n$ to be
equal to one. Here, the determinant of the operator-valued
$(n+1)\times (n+1)$ matrix is defined as $\det C=\sum_{\sigma\in
G_{n+1}} sgn(\sigma)C_{\sigma(1),1}C_{\sigma(2),2}\ldots
C_{\sigma(n+1),n+1}$, where $G_{n+1}$ is a set of all possible
permutations of the integers $\{1,\ldots,n+1\}$.

Particularly, when  $\psi_i$ are periodic functions except even
number of antiperiodic ones, the Wronskian $W$ is periodic. Since
the derivatives do not change the period of the functions, $W_i$ are
periodic as well. The formulas above then justify periodicity of the
operator $A_n$.

Finally, let us make a few comments on the superpartner
Hamiltonians $H$ and $\tilde{H}$ intertwined by operator $A_n$
(see (\ref{crum})) which annihilates a part of the physical states
of $H$. Let the potential of $H$ be smooth and the Wronskian $W$
computed on the kernel of $A_n$ be a nodeless function. Then the
potential of $\tilde{H}$ is smooth as well. The operator $A_n$ can
be used in reconstruction of the eigenstates $\tilde\psi$ of
$\tilde{H}$ corresponding to the eingenstates $\psi\neq \psi_i$ of
$H$ with the same eigenvalue,
$    \tilde H \tilde\psi =E\tilde\psi, \quad H \psi =E\psi, \quad
    \tilde{\psi}=A_n\psi.$
These wave functions $\tilde \psi$ can be also represented as

\begin{equation}
    \tilde \psi=A_n\psi=\frac{W(\psi _{1},\ldots,\psi_{n},\psi)}{W(\psi
    _{1},\ldots,\psi_{n})}.
\end{equation}

\noindent
This receipt fails in the reconstruction of the states $\tilde{\psi}_i$,
which correspond to the same eigenvalue as $\psi_i$,
$\tilde{H}\tilde{\psi}_i=E_i\tilde\psi_i,\ {H}{\psi}_i=E_i\psi_i$,
where $\psi_i$ is annihilated by $A_n$.
These functions $\tilde \psi_i$,
annihilated by $A_n^{\dagger}$, are given by

\begin{equation}
    \tilde
    \psi_i=\frac{W(\psi _{1},..,\hat \psi_i,..,\psi_{n},)}{W(\psi
    _{1},\ldots,\psi_{n})}, \; i=1,...,n, \quad A_n^{\dagger}\tilde
    \psi_i=0,
\end{equation}

\noindent where the entry below a symbol ``$\,\hat{}\,$"  is
omitted.

\setcounter{equation}{0}
\renewcommand{\theequation}{B.\arabic{equation}}

\section*{Appendix B}
\subsection*{Grading $\boldsymbol{\Gamma_*=\sigma_3}$}

In this case, which corresponds to the usual choice of the grading
operator, the non-diagonal supercharges $\mathcal{Q}_{\pm}$ are
fermionic operators, $\{\mathcal{Q}_{\pm},\sigma_3 \}=0$, whereas
the diagonal integral $\mathcal{Z}$ is a bosonic generator. Table
\ref{T1} represents the explicit identification of the bosonic and
fermionic generators, and corresponding polynomials appearing in
the (anti)commutation relations.

\begin{table}[h!]
\begin{center}
\begin{tabular}{|c|c|c|c|c|}
\hline \textbf{Fermionic } & $F_{1}=\mathcal{Q}_{-}$ &
$F_{2}=-R\sigma
_{3}\mathcal{Q}_{-}$ & $F_{3}=-iR\mathcal{Q}_{-}$ & $F_{4}=i\sigma _{3}\mathcal{Q}_{-}$ \\
\cline{2-5}
\multicolumn{1}{|c|}{\textbf{integrals}} & $F_{5}=R\mathcal{Q}_{+}$ & $F_{6}=-\mathcal{Q}_{+}$ & $%
F_{7}=i\sigma _{3}\mathcal{Q}_{+}$ & $F_{8}=-iR\sigma _{3}\mathcal{Q}_{+}$ \\
\hline\hline
\textbf{Bosonic}  & $\mathcal{H}$ & $\Sigma _{1}=-R$ & $\Gamma_* =\sigma _{3}$ & $%
\Sigma _{2}=-R\sigma _{3}$ \\ \cline{2-5}
\multicolumn{1}{|c|}{\textbf{integrals}} & $B_{1}=-iR\sigma _{3}\mathcal{Z}$ & $%
~B_{2}=-\sigma _{3}\mathcal{Z}$ & $~B_{3}=-iR\mathcal{Z}$ & $B_{4}=-%
\mathcal{Z}$ \\ \hline\hline
{\textbf{Polynomials}}   & $P_{22}=P_{2B}=P_{-}
(\mathcal{H})$ &$P_{11}=P_{1B}=P_{+}(\mathcal{H})$ & $P_{12}=1$  & $P_{B}=P_{Z}(\mathcal{H})$  \\
\cline{1-5}
\end{tabular}
\end{center}
\caption{Integrals of motion and structure polynomials, grading
$\Gamma_* =\sigma _{3}$.} \label{T1}
\end{table}

\subsection*{Grading $\boldsymbol{\Gamma_*=R}$}

For the choice  $\Gamma_*=R$,  the parity-odd diagonal,
$\mathcal{Z}$, and non-diagonal, $\mathcal{Q}_-$, integrals  are
fermionic supercharges. The non-diagonal parity-even integral
$\mathcal{Q}_+$ is identified as a bosonic generator. The
identification of all the generators and structure polynomials are
given by Table \ref{T2}.

\begin{table}[h!]
\begin{center}
\begin{tabular}{|c|c|c|c|c|}
\hline
\textbf{Fermionic } & $F_{1}=\mathcal{Q}_{-}$ & $F_{2}=R\sigma _{3}\mathcal{Q}_{-}$ & $%
F_{3}=i\sigma _{3}\mathcal{Q}_{-}$ & $F_{4}=iR\mathcal{Q}_{-}$ \\
\cline{2-5} \textbf{integrals} & $F_{5}=iR\mathcal{Z}$ &
$F_{6}=iR\sigma _{3}\mathcal{Z}$ & $F_{7}=\sigma _{3}\mathcal{Z}$
& $F_{8}=\mathcal{Z}$
\\ \hline\hline
\textbf{Bosonic}  & $\mathcal{H}$ & $\Gamma_* =R$ & $\Sigma _{1}=\sigma _{3}$ & $%
\Sigma _{2}=R\sigma _{3}$ \\ \cline{2-5}
\multicolumn{1}{|c|}{\textbf{integrals}} & $B_{1}=\mathcal{Q}_{+}$ & $%
B_{2}=i\sigma _{3}\mathcal{Q}_{+}$ & $B_{3}=R\mathcal{Q}_{+}$ &
$B_{4}=iR\sigma _{3}\mathcal{Q}_{+}$
\\ \hline\hline
{\textbf{Polynomials}}   & $P_{22}=P_{12}=P_{-}
(\mathcal{H})$ &$P_{11}=P_{Z}(\mathcal{H})$ & $P_{2B}=1$  & $P_{B}=P_{1B}=P_{+}(\mathcal{H})$  \\
\cline{1-5}
\end{tabular}
\end{center} \caption{Integrals of motion and structure polynomials,
grading $\Gamma_* =R$.}\label{T2}
\end{table}

\subsection*{Grading $\boldsymbol{\Gamma_*=R\sigma_3}$}

With this choice of the grading operator, integrals  $\mathcal{Z}$
and $\mathcal{Q}_+$ are identified as fermionic supercharges,
integral  $\mathcal{Q}_-$ is a bosonic generator. Complete
identification of the generators and structure polynomials are
represented by Table \ref{T3}.

\begin{table}[h!]
\begin{center}
\begin{tabular}{|c|c|c|c|c|}
\hline
\textbf{Fermionic } & $F_{1}=\mathcal{Z}$ & $F_{2}=-\sigma _{3}\mathcal{Z}$ & $%
F_{3}=-iR\mathcal{Z}$ & $~F_{4}=iR\sigma _{3}\mathcal{Z}$ \\
\cline{2-5}
\multicolumn{1}{|c|}{\textbf{integrals}} & $F_{5}=R\mathcal{Q}_{+}$ & $~F_{6}=-\mathcal{Q}_{+}$ & $%
F_{7}=iR\sigma _{3}\mathcal{Q}_{+}$ & $F_{8}=-i\sigma _{3}\mathcal{Q}_{+}$ \\
\hline\hline
\textbf{Bosonic}  & $\mathcal{H}$ & $\Sigma _{1}=-R$ & $\Sigma _{2}=-\sigma _{3}$ & $%
\Gamma_* =R\sigma _{3}$ \\ \cline{2-5}
\multicolumn{1}{|c|}{\textbf{integrals}} & $B_{1}=-i\sigma _{3}\mathcal{Q}_{-}$ & $%
~~B_{2}=-R\sigma _{3}\mathcal{Q}_{-}$ & $~B_{3}=-iR\mathcal{Q}_{-}$ & $B_{4}=-\mathcal{Q}_{-}$ \\
\hline\hline
{\textbf{Polynomials}}   & $P_{22}=P_{Z}(\mathcal{H})$ &$P_{11}=
P_{12}=P_{+}(\mathcal{H})$ & $P_{1B}=1$  & $P_{B}=P_{2B}=P_{-}(\mathcal{H})$  \\
\cline{1-5}
\end{tabular}
\end{center}
\caption{Integrals of motion and structure polynomials, grading
$\Gamma_* =R\protect\sigma _{3}$.} \label{T3}
\end{table}

The anti-commutation relations between the fermionic operators are
given in Table \ref{T4}, while Table \ref{T5} provides the
boson-fermion commutation relations.

\begin{table}[h!]
\centering%
\begin{tabular}{|c||c|c|c|c|c|c|c|c|}
\hline & $F_{1}$ & $F_{2}$ & $F_{3}$ & $F_{4}$ & $F_{5}$ &
$F_{6}$ & $F_{7}$ & $F_{8}$ \\
\hline\hline
$F_{1}$ & $P_{22}$ & $\Sigma _{2}P_{22}$ & $0$ & $0$ & $0$ & $B_{4}P_{12}$ & $%
0$ & $B_{1}P_{12}$ \\ \hline $F_{2}$ & $\Sigma _{2}P_{22}$ &
$P_{22}$ & $0$ & $0$ &
$-B_{2}P_{12}$ & $0$ & $-B_{3}P_{12}$ & $0$ \\
\hline $F_{3}$ & $0$ & $0$ & $P_{22}$ & $\Sigma
_{2}P_{2}$ & $0$ & $-B_{3}P_{12}$ & $0$ & $B_{2}P_{12}$ \\
\hline
$F_{4}$ & $0$ & $0$ & $\Sigma _{2}P_{22}$ & $P_{22}$ & $B_{1}P_{12}$ & $0$ & $%
-B_{4}P_{12}$ & $0$ \\ \hline
$F_{5}$ & $0$ & $-B_{2}P_{12}$ & $0$ & $B_{1}P_{12}$ & $%
P_{11}$ & $\Sigma _{1}P_{11}$ & $0$ & $0$ \\
\hline
$F_{6}$ & $B_{4}P_{12}$ & $0$ & $-B_{3}P_{12}$ & $0$ & $%
\Sigma _{1}P_{11}$ & $P_{11}$ & $0$ & $0$ \\
\hline
$F_{7}$ & $0$ & $-B_{3}P_{12}$ & $0$ & $-B_{4}P_{12}$ & $0$ & $%
0$ & $P_{11}$ & $\Sigma _{1}P_{11}$ \\ \hline $F_{8}$ &
$B_{1}P_{12}$ & $0$ & $B_{2}P_{12}$ & $0$ &
$0$ & $0$ & $\Sigma _{1}P_{11}$ & $P_{11}$ \\
\hline
\end{tabular}
\caption{Fermion-fermion anti-commutation relations. Here the
overall multiplicative factor~$2$ is omitted. To get
anti-commutator, the corresponding entry should be multiplied
by~$2$, for instance, $\{F_1,F_1\}=2P_{22}$.}\label{T4}
\end{table}

\begin{table}[h!]
\centering%
\begin{tabular}{|c||c|c|c|c|c|c|c|c|}
\hline & $F_{1}$ & $F_{2}$ & $F_{3}$ & $F_{4}$ & $F_{5}$ &
$F_{6}$ & $F_{7}$ & $F_{8}$ \\
\hline\hline
$\Gamma_*$ & $-iF_{4}$ & $-iF_{3}$ & $iF_{2}$ & $iF_{1}$ & $iF_{8}$ & $%
iF_{7}$ & $-iF_{6}$ & $-iF_{5}$ \\ \hline $\Sigma _{1}$ &
$-iF_{3}$ & $-iF_{4}$ & $iF_{1}$ & $iF_{2}$ & $0$ & $0$ & $0$ &
$0$ \\ \hline $\Sigma _{2}$ & $0$ & $0$ & $0$ & $0$ & $iF_{7}$ &
$iF_{8}$ & $-iF_{5}$ & $-iF_{6}$ \\ \hline $B_{1}$ & $0$ &
$iF_{6}P_{2B}$ & $-iF_{7}P_{2B}$ & $0$ & $0 $ & $-iF_{2}P_{1B}$ &
$iF_{3}P_{1B}$ & $0$ \\ \hline $B_{2}$ & $iF_{7}P_{2B}$ & $0$ &
$0$ &
$iF_{6}P_{2B}$ & $0$ & $-iF_{4}P_{1B}$ & $-iF_{1}P_{1B}$ & $0$ \\
\hline
$B_{3}$ & $-iF_{5}P_{2B}$ & $0$ & $0$ & $iF_{8}P_{2B}$ & $%
iF_{1}P_{1B}$ & $0$ & $0$ & $-iF_{4}P_{1B}$ \\ \hline
$B_{4}$ & $0$ & $-iF_{8}P_{2B}$ & $-iF_{5}P_{2B}$ & $0$ & $%
iF_{3}P_{1B}$ &$0$ & $0$ & $iF_{2}P_{1B}$ \\
\hline
\end{tabular}
\caption{Boson-fermion commutation relations. The  overall
multiplicative factor $2$ is omitted. To get commutator, the
corresponding entry should be multiplied by~$2$, for instance,
$[\Gamma_*,F_1]=-2iF_{4}$.}\label{T5}
\end{table}

\setcounter{equation}{0}
\renewcommand{\theequation}{C.\arabic{equation}}

\section*{Appendix C}

In the treatment of the section $4.1$, we left untouched the
system described by Lam\'e associated Hamiltonian with $m=l$. Due
to an identity $\mathrm{dn}\, (x+K)=k'/\mathrm{dn}\, x$, in
contrary to the other members of the family, its period is $K$.
This fact explains why the algebraic methods applied to other
members of the family in this case say that the dimension of
$sl(2,\R)$ representation realized on antiperiodic in the period
$2K$ singlet states is equal to $m-l=0$. This is just because
singlet states with such a period  does not exist. The place of
this system in the mosaic of the tri-supersymmetric
self-isospectral systems is clarified by its intimate relation to
pure Lam\'e system, mediated by Landen transformation
\cite{landen,SukhKhLan}.

Landen's transformation of the elliptic functions can be written as
\begin{eqnarray}
    \mathrm{sn}\ (x,k)&=&\alpha\frac{\mathrm{sn}\left(\frac{x}{\alpha},\kappa
    \right)\mathrm{cn} \left(\frac{x}{\alpha},
    \kappa\right)}{\mathrm{dn}\left(\frac{x}{\alpha},\kappa\right)},\ \
    \mathrm{cn}\ (x,k)=\frac{1-{\alpha}\ \mathrm{sn}^2(\frac{x}{\alpha},\kappa)}
    {\mathrm{dn}\ (\frac{x}{\alpha},\kappa)},\ \nonumber\\
    \mathrm{dn}\ (x,k)&=&\frac{\kappa'+(2-\alpha)\
    \mathrm{cn}^2(\frac{x}{\alpha},\kappa)}{\mathrm{dn}\ (\frac{x}{\alpha},
    \kappa)}\label{landentr}
\end{eqnarray}
where $\alpha=\frac{2}{1+k},$  $\kappa^2=\frac{4k}{(1+k)^2},$
$k=\frac{1-\kappa'}{1+\kappa'},$ $\alpha=1+\kappa'$. To avoid
confusions, let us denote explicitly the dependence of the
complete elliptic integral $K$ on the modular parameter such that
we will write $K(k)$ or $K(\kappa)$. Since $K(\kappa)=(1+k)K(k)$,
Landen's transformation divides in two the period $2K(k)$ of the
elliptic functions in the sense that the period of the resulting
expression is $K(\kappa)$.

Using the identities (\ref{landentr}), we can rewrite the  Lam\'e
Hamiltonian in terms of the elliptic functions of a new variable
$y=\frac{x}{\alpha}$ and the modular parameter $\kappa$,
 $   H_{m,0}^-(x,k)
     =\frac{1}{\alpha^2}\left[H_{m,m}(y,\kappa)\right]+const.$
The displacement $K(k)$ of the pure Lam\'e
tri-supersymmetric partner  changes to $\frac{K(\kappa)}{2}$ in the
case $m=l$. It is in accordance with our result on the general case
$m\neq l$ where the superpartner potential was displaced in the half
of the real period. Thus, we obtain finally the relation
 \begin{eqnarray}
    \left(\begin{array}{cc}H_{m,0}^-(x+K(k),k)&0\\
    0& H_{m,0}^-(x,k)
    \end{array}
    \right)=\frac{1}{\alpha^2}\left(\begin{array}{cc}
    H_{m,m}^-\left(y+\frac{K(\kappa)}{2},\kappa\right)&0\\
    0& H_{m,m}^-(y,\kappa)
    \end{array}
    \right)+c,\label{lame=m}
\end{eqnarray}
\noindent where $c$ is a constant term.

It suggests directly the form of the tri-supersymmetry in the
special case of $m=l$ Lam\'e  associated systems; all the operators
commuting with $\mathcal{H}_{m,0}$ commute with $\mathcal{H}_{m,m}$
as well. To get their explicit form for the systems described by
$\mathcal{H}_{m,m}$ we just have to rescale the variable and apply
the identities (\ref{landentr}) in the formulas (\ref{matrixHXY})
for $\mathcal{X}_{m,0}(x,k),\ \mathcal{Y}_{m,0}(x,k)$ and
$\mathcal{Z}_{m,0}(x,k)$. Then we can write immediately
 \begin{eqnarray}
     \mathcal{X}_{m,m}(y,\kappa)&=&\mathcal{X}_{m,0}
     \left(\alpha y,k(\kappa)\right)=\mathcal{X}_{m,0}
     \left((1+\kappa') y,\frac{1-\kappa'}{1+\kappa'}\right),\nonumber\\
     \mathcal{Y}_{m,m}(y,\kappa)&=&\mathcal{Y}_{m,0}
     \left(\alpha y,k(\kappa)\right)=\mathcal{Y}_{m,0}
    \left((1+\kappa') y,\frac{1-\kappa'}{1+\kappa'}\right),
    \nonumber\\ \mathcal{Z}_{m,m}(y,\kappa)&=&\mathcal{Z}_{m,0}
     \left(\alpha y,k(\kappa)\right)=\mathcal{Z}_{m,0}
     \left((1+\kappa') y,\frac{1-\kappa'}{1+\kappa'}\right). \label{XYZm=l}
 \end{eqnarray}

\noindent Naturally, the algebraic relations between the operators
remain unchanged. Thus, the self-isospectral supersymmetry and
associated superalgebra are recovered for $m=l$ case.

\newpage

\end{document}